\documentstyle[epsfig,cite,12pt]{article}
\newcommand{\etal}{{\it et al.}}

\newcommand{\bu}{\(b \to \rm{X}_{\rm{u}} \ell \nu\ \)}
\newcommand{\bc}{\(b \to \rm{X}_{\rm{c}} \ell \nu\ \)}
\newcommand{\ub}{\(|\rm{V}_{\rm{ub}}|\)\ }
\newcommand{\uub}{|\rm{V}_{\rm{ub}}|}
\newcommand{\brbu}{Br(\(b \to \rm{X}_{\rm{u}} \ell \nu\))\ }
\newcommand{\brbr}{\rm{Br}(\it{b} \to \rm{X}_{\rm{u}} \ell \nu)}

\begin{document}
\bibliographystyle{plain}
\begin{titlepage}
\begin{center}{\large   EUROPEAN ORGANIZATION FOR NUCLEAR RESEARCH
 }\end{center}\bigskip
\begin{flushright}
CERN-EP-2001-044 \\
June 28, 2001  \\
\end{flushright}
\bigskip\bigskip\bigskip\bigskip\bigskip
\begin{center}
 {\huge\bf \boldmath
Measurement of  $|\rm{V}_{\rm{ub}}|$  using $b$ hadron semileptonic  
decay
}
\end{center}
\bigskip\bigskip
\begin{center}{\LARGE The OPAL Collaboration}
\end{center}\bigskip\bigskip
\bigskip\begin{center}{\large  Abstract}\end{center}

The magnitude of the CKM matrix element \ub is determined by 
measuring the inclusive charmless
semileptonic branching fraction of beauty hadrons at
OPAL based on \bu event topology and kinematics.
This analysis uses OPAL data collected between 1991 and 1995,
which correspond to  about four million hadronic Z decays.
We measure \brbu to be
(1.63 $\pm$ 0.53 $  ^{+0.55}_{-0.62}$) $\times$ $10^{-3}$. The
 first uncertainty
is the statistical error and the second is the systematic error.
From this analysis, \ub is determined to be:\\
\[ |\rm{V}_{\rm{ub}}| = (4.00 \pm 0.65\ (\rm{stat}) \
^{+0.67}_{-0.76}\ (\rm{sys}) \pm 0.19\ (\rm{HQE})) \times  10^{-3}. 
\]\\
The last error represents the theoretical uncertainties related to the 
extraction of \ub from \brbu using the Heavy Quark Expansion.

 \bigskip
\begin{center}

 \bigskip
 \bigskip

 \end{center}
\begin{center}
{\large (To be submitted to the Eur. Phys. J. C)}
 \end{center}
\end{titlepage}


\begin{center}{\Large        The OPAL Collaboration
}\end{center}\bigskip
\begin{center}{
G.\thinspace Abbiendi$^{  2}$,  
C.\thinspace Ainsley$^{  5}$,
P.F.\thinspace {\AA}kesson$^{  3}$,
G.\thinspace Alexander$^{ 22}$,
J.\thinspace Allison$^{ 16}$,
G.\thinspace Anagnostou$^{  1}$,
K.J.\thinspace Anderson$^{  9}$,
S.\thinspace Arcelli$^{ 17}$,
S.\thinspace Asai$^{ 23}$,
D.\thinspace Axen$^{ 27}$
% ...
}\end{center}\bigskip
\bigskip
$^{  1}$School of Physics and Astronomy, University of Birmingham,
Birmingham B15 2TT, UK
\newline
$^{  2}$Dipartimento di Fisica dell' Universit\`a di Bologna and INFN,
I-40126 Bologna, Italy
\newline
$^{  3}$Physikalisches Institut, Universit\"at Bonn,
D-53115 Bonn, Germany
\newline
% ...
$^{  5}$Cavendish Laboratory, Cambridge CB3 0HE, UK
\newline
% ...
$^{  9}$Enrico Fermi Institute and Department of Physics,
University of Chicago, Chicago IL 60637, USA
\newline
% ...
$^{ 16}$Department of Physics, Schuster Laboratory, The University,
Manchester M13 9PL, UK
\newline
$^{ 17}$Department of Physics, University of Maryland,
College Park, MD 20742, USA
\newline
% ...
$^{ 22}$Department of Physics and Astronomy, Tel Aviv University,
Tel Aviv 69978, Israel
\newline
$^{ 23}$International Centre for Elementary Particle Physics and
Department of Physics, University of Tokyo, Tokyo 113-0033, and
Kobe University, Kobe 657-8501, Japan
\newline
% ...
$^{ 27}$University of British Columbia, Department of Physics,
Vancouver BC V6T 1Z1, Canada
\newline
% ...
Tel Aviv 69978, Israel.

\newpage


\section{Introduction}
 \label{sec:intro}
The  CKM matrix \cite{CKM}    describes
the relation between quark weak and mass eigenstates, with the 
element $\rm{V}_{\rm{ub}}$ describing decays of the b to u quark.
Its magnitude,
\(|\rm{V}_{\rm{ub}}|\),  can be calculated
by measuring the inclusive  \(\rm{b} \to \rm{u}\) semileptonic
decay rate. Given that the branching fraction of inclusive  
\( \rm{b} \to \rm{u}\) semileptonic
decay is  of order $10^{-3}$,  a large number of $b$ hadrons 
are required to measure  \(|\rm{V}_{\rm{ub}}|\).
The dominant background to \bu 
comes  from \bc decays 
because  the branching ratio
of \bc is more than 50 times greater than that of \(b \to
\rm{X}_{\rm{u}} \ell \nu \).
Here the lepton  $\ell$ refers to either an electron or a muon, and $b$
denotes all weakly decaying $b$ hadrons$^($\footnote{Charged conjugate states
are implied if not stated otherwise.\label{fn:1}}$^)$. 
$\rm{X}_{\rm{u}}$ and $\rm{X}_{\rm{c}}$ represent 
hadronic states  resulting from a b quark  semileptonic
decay to a u  or c quark  respectively. The determination 
 of \ub depends on  the b to u and b  to c
semileptonic decay models.

The  inclusive method developed by ARGUS\cite{ARGUS} and
CLEO\cite{CLEO4} is to extract
$|\rm{V}_{\rm{ub}}|/|\rm{V}_{\rm{cb}}|$
from the excess of  events in  the  2.3 to 2.6 GeV/$c$ region of
the lepton momentum spectrum in the B meson  rest frame, 
where the \bc contributions vanish. This technique uses only a small 
fraction of the lepton phase space and so has considerable model 
dependence in extrapolating to the entire lepton spectrum in the B 
rest frame.
In addition, since the LEP experiments can not precisely determine the B meson rest frame,
this method is not appropriate for the LEP experiments.
Instead, at LEP,  \ub or $|\rm{V}_{\rm{ub}}|/|\rm{V}_{\rm{cb}}|$  is extracted
using  a larger portion of the  lepton spectrum as well as 
other
kinematic variables. The inclusive measurement
of the branching  fraction of the \bu decay has
been performed at LEP  by  ALEPH\cite{ALEPH},
DELPHI\cite{DELPHI} and  L3\cite{L3}.

The  theoretical uncertainty for the value of \ub 
extracted from a measurement of inclusive
\bu branching fraction differs  from that 
extracted from measurements of exclusive
b $\to$ u semileptonic decay rates. 
A recent theoretical study concludes that there is a  5$\%$
theoretical uncertainty on \ub values derived from
\bu inclusive measurements\cite{HQE}, using the Heavy Quark
Expansion.  There is a  15$\%$  theoretical
uncertainty  associated with  \ub values extracted from measurements 
of the exclusive branching fractions  \(\rm{B} \to \pi \ell\nu\) or 
\(\rm{B} \to \rho \ell\nu\) \cite{CLEO99}, interpreted within the
framework of the Heavy Quark Effective Theory (HQET).

In this paper, we describe  the  determination of \ub using 
the inclusive \bu decay rate from  the OPAL data taken at center of 
mass energies near the Z resonance.
The event preselection, the \bu decay models and
the neural network  used to separate \bu from the background
will be discussed in detail in the following sections.

\section{The OPAL detector, data and Monte Carlo samples}
\label{sec:detector}
The OPAL detector is a multi-purpose 4$\pi$  spectrometer incorporating
  excellent charged and neutral particle detection capabilities.
The OPAL detector is described  in detail  elsewhere\cite{detector}. 
A brief description is given here. The central tracking
system consists of a silicon microvertex detector, a vertex chamber, 
a jet chamber and $z$ chambers. The momentum of tracks and  the primary
and secondary vertex position   are  reconstructed
by the central tracking system, which is located inside a solenoid.
The solenoid  provides  a magnetic field of 0.435T. Outside the 
solenoid  is the electromagnetic calorimeter, which is  composed of 
lead glass blocks and is used to measure
the energies and positions of electrons and photons. The hadron calorimeter lies
outside the electromagnetic calorimeter and is used to measure the energy of hadrons 
emerging from the electromagnetic calorimeter and assists in the
identification of muons. The outermost  OPAL detector is the
muon detector which  consists of a system of   barrel and endcap muon 
chambers. A large fraction of muons with momenta less than
2 GeV/$c$ are absorbed by the other detectors or the iron shielding 
before reaching the muon chambers.

The current analysis uses OPAL 1991 to 1995 data, collected near the
Z resonance, comprising about  four  million  hadronic  Z   decays.
Monte Carlo simulated events were generated using the JETSET 7.4\cite{JETSET}
generator, with  parameters described in \cite{para}.
Approximately five million hadronic  Z $\to$ b$\bar{\rm{b}}$ decays 
were generated to study the \bc decay
and the \( \rm{b} \to \rm{c} \to \ell\) cascade decay. 
Six  million hadronic \(\rm{Z} \to \rm{q}\bar{\rm{q}}\)
(where q can be u, d, s, c and b) decays
were generated to study the leptons from primary charm quarks
and light quarks. Two hundred thousand events from a
\bu hybrid model\cite{hybrid} were produced to simulate
the \( \rm{b} \to\rm{u} \)
semileptonic decay. The hybrid model will be described in detail
in Section~\ref{sec:method}.

\section{Signal and background simulation}
   The b to u semileptonic decay  and  background simulation are
described below. The b to u semileptonic decay and
background simulated events are passed through  the full OPAL 
detector simulation \cite{gopal} to produce the  corresponding response.
For this paper, the  production fractions of $\rm{B}^{+}$, 
$\rm{B}^0$, $\rm{B}^0_{\rm{s}}$ and  $\Lambda_{\rm{b}}$ 
in Z decays   were  adjusted to reproduce those given by the Particle
 Data Group\cite{PDG}.
\subsection{\boldmath The \bu hybrid model}
\label{sec:method}
Several theoretical models have been proposed for the 
\bu decay. Exclusive bound-state models
\cite{bound1,IGSW,bound2, ISGW2} 
approximate the inclusive \bu lepton  spectra  by
summing contributions from all the exclusive final states. 
These exclusive  models do  not include  all the  possible 
final states  nor any non-resonant states and therefore yield an 
incomplete prediction of the inclusive
lepton momentum distribution, especially in the region of  high 
hadronic invariant mass. The inclusive free 
quark models\cite{ACCMM,QCD,QCD1,QCD3,parton}
treat the heavy quark  as a free quark and
the final state  as a  quark plus gluons. 
Free quark models are known to give  poor agreement with 
experiments at low u quark  recoil momentum.
Therefore,  a hybrid model\cite{hybrid} has been proposed 
to  model the  \bu decay by using the exclusive model in the lower 
hadronic invariant mass region and using the inclusive model in the 
higher hadronic invariant mass region.
The ISGW2  model \cite{ISGW2}  is used as the exclusive part 
of the hybrid model. The ACCMM model \cite{ACCMM}, combined
with the W  decay model \cite{wdecay} plus JETSET fragmentation, 
is used as the inclusive part of the hybrid model.
Since the  ISGW2 exclusive model includes the exclusive resonant final
states   1S, 2S and  1P  up to 1.5 GeV/$c^2$  in the hadronic mass,
the boundary between the inclusive and exclusive parts of the hybrid
model is placed at the hadronic invariant mass  of 1.5 GeV/$c^2$.
The relative normalization of the inclusive and exclusive parts of
the hybrid model is determined by the inclusive model. 
This hybrid model is only applied to decays of B mesons. There are
no theoretical predictions for  b to u semileptonic transitions
of $b$ baryons. The exclusive transitions of the $b$ baryons in 
the OPAL tune of  JETSET\cite{JETSET,para} are used. 

\begin{figure}[htbp]
\centering
\epsfig{file=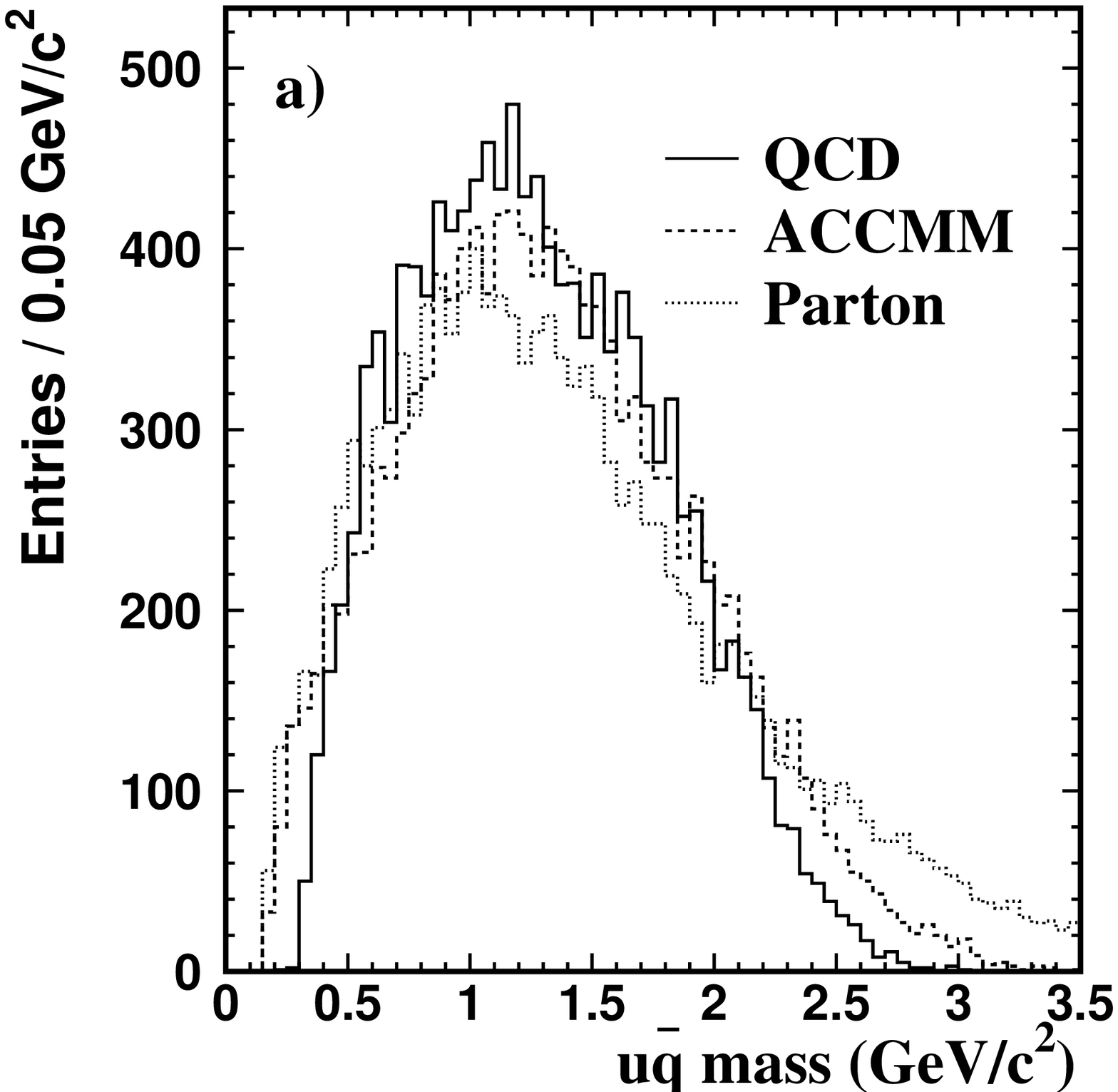,width=3. in}
\epsfig{file=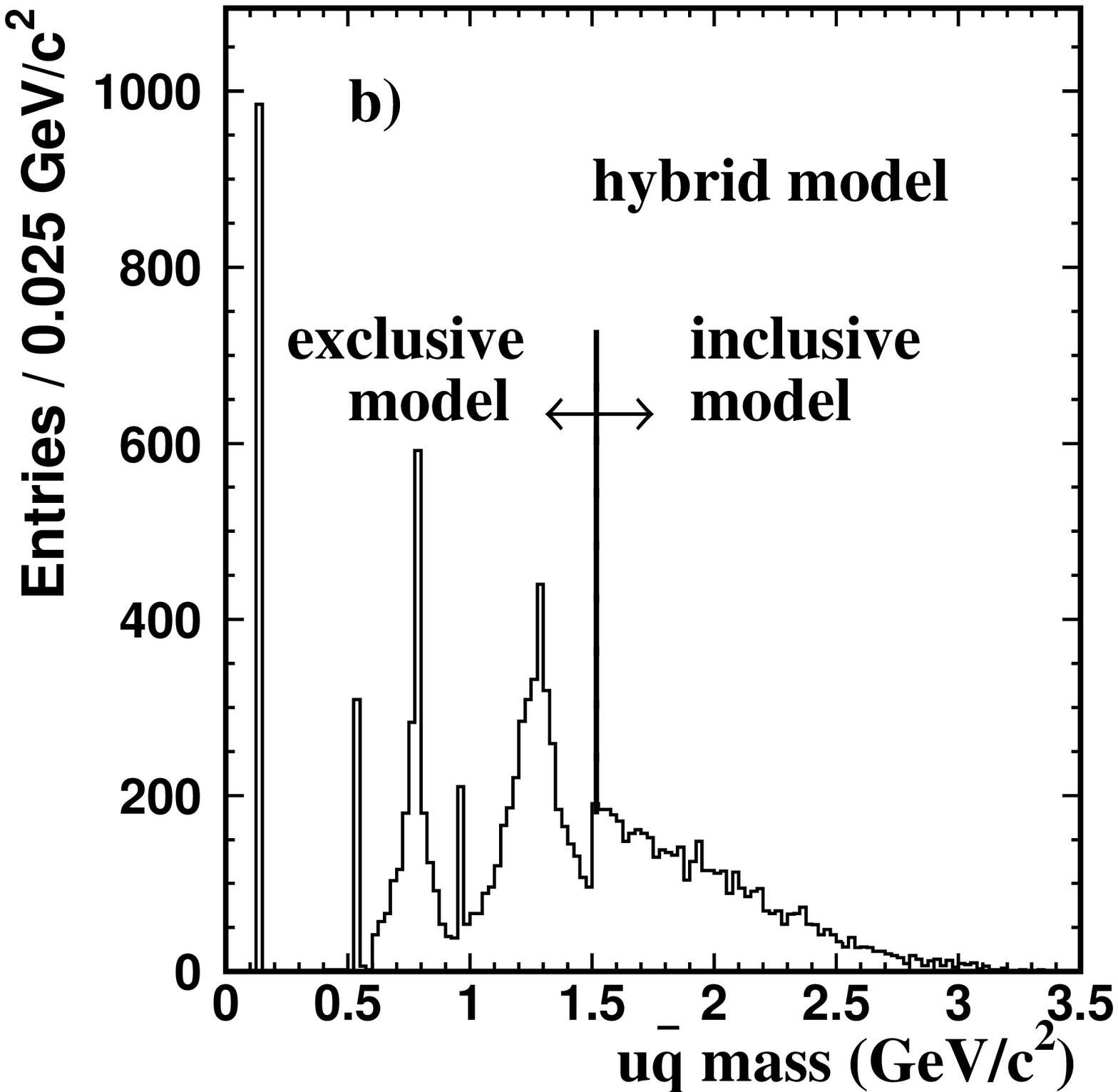,width=3. in}
\caption{\textbf{a,b.} The $\rm{u}\bar{\rm{q}}$ invariant mass distributions,
\textbf{a} using the QCD universal function, ACCMM and parton inclusive  models,
\textbf{b} using the  hybrid model. Only the portion of the
$\rm{u}\bar{\rm{q}}$ invariant mass 
above 1.5 GeV/$c^2$ from the inclusive model in \textbf{a} is used in
the hybrid model. The boundary between the exclusive model (left arrow) and the
inclusive model (right arrow) in the hybrid model is indicated by the
dashed line  in \textbf{b}.}
\label{fig:3models}
\end{figure}

In  order to  estimate systematic uncertainties due to modeling of
the inclusive spectrum, alternative models are also studied.
 Signal events were generated  with the QCD universal
 function \cite{QCD,QCD1,QCD3} and parton\cite{parton} models.
The invariant mass distributions of the hadronic recoil 
$\rm{u}\bar{\rm{q}}$ system  are shown 
in Figure~\ref{fig:3models}a for the QCD universal function, 
ACCMM and parton models.  The invariant mass distribution of the hadronic recoil
$\rm{u}\bar{\rm{q}}$  system for the hybrid model is  shown in
Figure~\ref{fig:3models}b.

\subsection{Background simulation}
The ACCMM model\cite{ACCMM} is used to describe the lepton spectrum of
 \bc and  \( \rm{b} \to \rm{c} \to \ell \) decays. The fragmentation
function of Peterson \etal \cite{Peterson} is used to describe the b
quark and c quark fragmentation. The branching fractions of
\( \rm{B}^{0} \to \rm{D}^{-}\ell^{+} \nu \), \( \rm{B}^{0} \to 
\rm{D}^{*-}\ell^{+} \nu \),
\( \rm{B}^{+} \to \bar{\rm{D}}^{0}\ell^{+} \nu \), \( \rm{B}^{+} \to 
\bar{\rm{D}}^{*0}\ell^{+} \nu \) and \( \Lambda_{\rm{b}} \to \Lambda_{\rm{c}}\rm{X}\ell\nu \)
 were modified to reproduce those given by the Particle
Data Group \cite{PDG}. The $\Lambda_{\rm{b}}$ lepton
momentum spectrum corresponding to  -56$\%$  polarization  
\cite{polar} was used.

\section{Event preselection}
\label{sec:evsel}
A hadronic event selection \cite{electron} and detector
performance requirements are applied to  the data.
The thrust polar angle$^($\footnote{ A right handed coordinate system is
used, with positive $z$  along the $\rm{e}^{-}$  beam direction and
$x$  pointing toward the center of the LEP ring.
 The polar and azimuthal angles are denoted by $\theta$  and $\phi$,  and the
 origin is taken to be the center of the detector.\label{fn:2}}$^)$
 $|\cos\theta|$ is required to be less than 0.9 to ensure that the
events are well contained within the acceptance of the detector.
The selected events must pass the   $b$  identification, the lepton selection
and the $b$  semileptonic decay selection. All these selections are
described in detail in the following sections. After all these
preselections, the \bu decay purity is 1.3$\%$  and the main background is 
from  \bc decays.
\subsection{\boldmath $b$  identification}
A neural network algorithm \cite{rb} based on charged particle vertex
information is used to separate  the b flavour events from the  other flavour
events in each hemisphere.  If either hemisphere passes this neural network selection,
the event is selected. After this neural network selection,
the  $b$ purity is  more than 91$\%$ and the  $b$ identification
selection  efficiency is approximately 30$\%$ per hemisphere from the
Monte Carlo simulation in which a branching fraction of
$1.0 \times 10^{-3}$ for the $ b \to X_u \ell \nu$ transition is incorporated.
Both hemispheres are searched for  electron and muon candidates after
the $b$ identification. 
\subsection{Lepton selection}
Electrons are identified by a neural network\cite{rb} using
the track and calorimeter information.
The electron momentum is required to be greater than  2
GeV/$c$. Electrons from  photon
conversions, \( \gamma \to \rm{e}^{+}\rm{e}^{-} \), contribute a 
significant background to the prompt electron samples. 
Another neural network is used to reject this background \cite{rb}. 
The photon conversion background is reduced by 94$\%$  after the photon 
conversion neural network selection, whilst retaining 98$\%$  of the selected
prompt electrons. After all these requirements, the resulting electron efficiency
is approximately 74$\%$  with a purity of 94$\%$ within the 
geometrical acceptance.

Muons are identified  using reconstructed track segments in the muon
chambers\cite{rb}.  The muon momentum is required to be greater than  3  GeV/$c$.
The reconstructed tracks in the central detector are
extrapolated to the muon chambers to see if they match the track
segments reconstructed in the external muon chambers. The measured
energy loss dE/dx is also required to be consistent with the expected
value for a  muon. After all these requirements,
the  muon selection efficiency is approximately 90$\%$ and
the muon purity  approximately 93$\%$ within the 
geometrical acceptance.

Electron and muon  momenta transverse to the direction of the jet
containing the lepton 
are required to be greater than 0.5 GeV/$c$ in order to reject  
leptons from light quark decay. The lepton is included in the calculation of
the jet direction. The jet finding is based on the cone 
algorithm\cite{cone}.
\subsection{\boldmath $b$ semileptonic decay selection}
\begin{figure}[htbp]
\epsfig{file=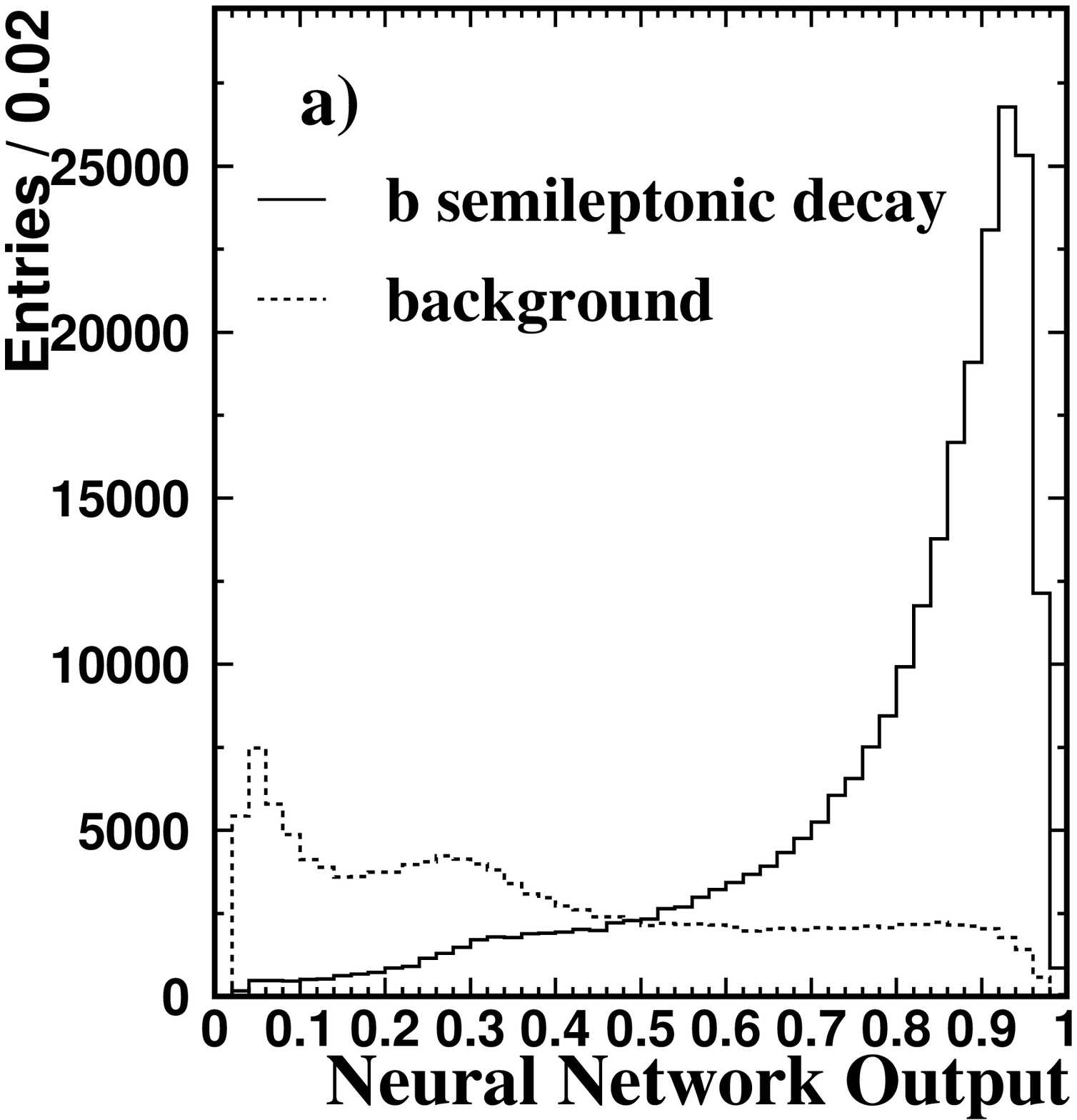,width=3. in}
 \epsfig{file=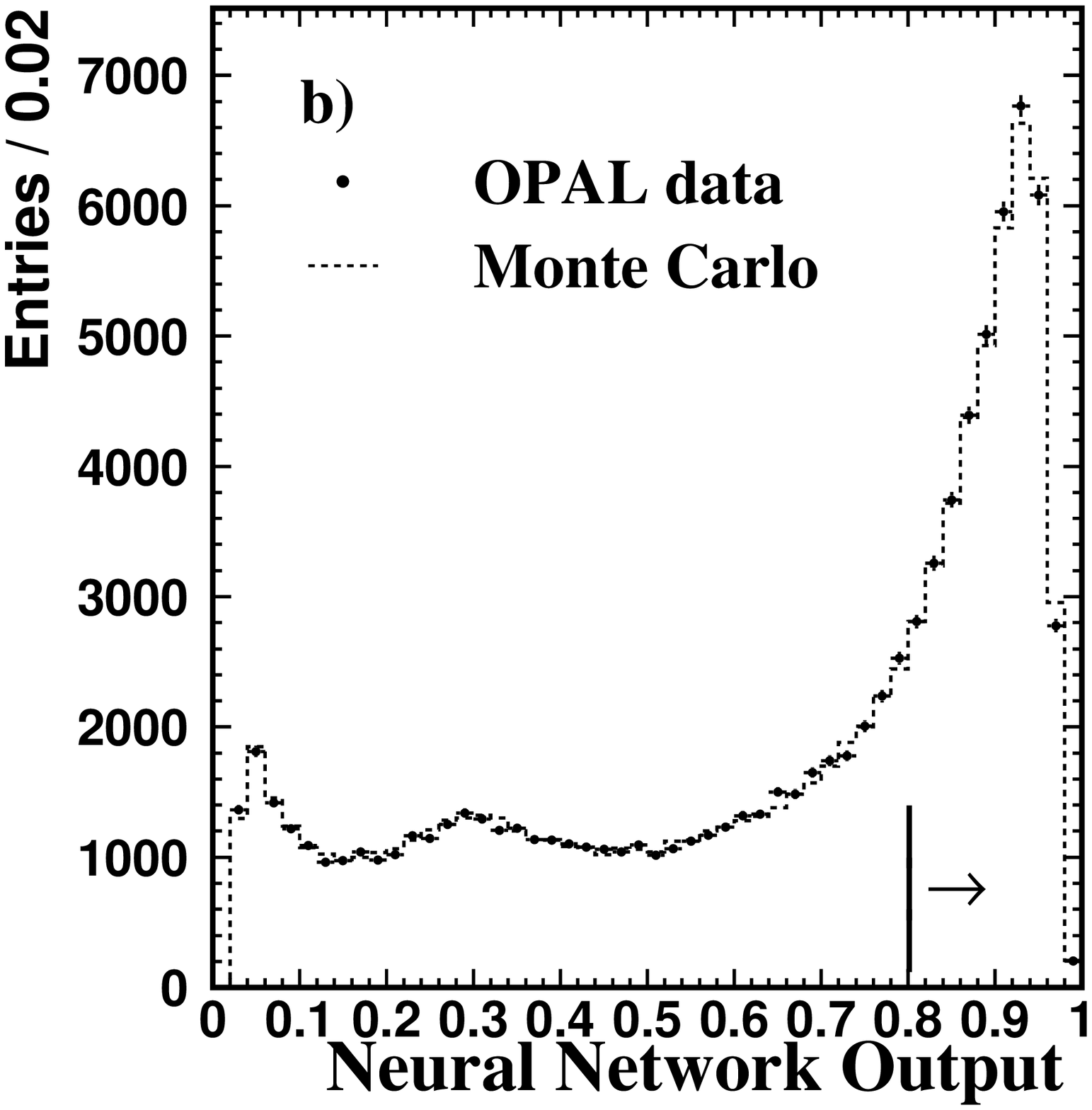,width=3. in}
\caption{\textbf{a,b.} The  $b$ hadron semileptonic decay neural 
network output distributions.
\textbf{a}  for
the $b$  semileptonic decays and the scaled background 
from the Monte Carlo simulated events,
here the background indicates all
events excluding \bc and \bu,
\textbf{b} comparison
between the OPAL data and the Monte Carlo simulated events.
The selected region is shown by the  arrow in \textbf{b}.
}
\label{fig:nnbl}
\end{figure}

 A neural network \cite{chris} based on lepton information is used
 to separate the $b$ hadron  semileptonic decays,
\bc and \(b \to \rm{X}_{\rm{u}} \ell \nu\), from 
non $b$ semileptonic decays.  The distributions of the  neural network 
output variable are shown in Figure~\ref{fig:nnbl}. After this neural 
network $b$ semileptonic decay selection,
the $b$ hadron semileptonic decay purity is 97$\%$ and
 the efficiency is 65$\%$ for this neural network;
the c $\to$ $\ell$ events, where c is a primary quark,
and \mbox{b $\to$ c $\to$ $\ell$} events are suppressed.

\section{\boldmath \bu neural network}
\label{sec:5}
\begin{figure}[htbp]
\centering
\epsfig{file=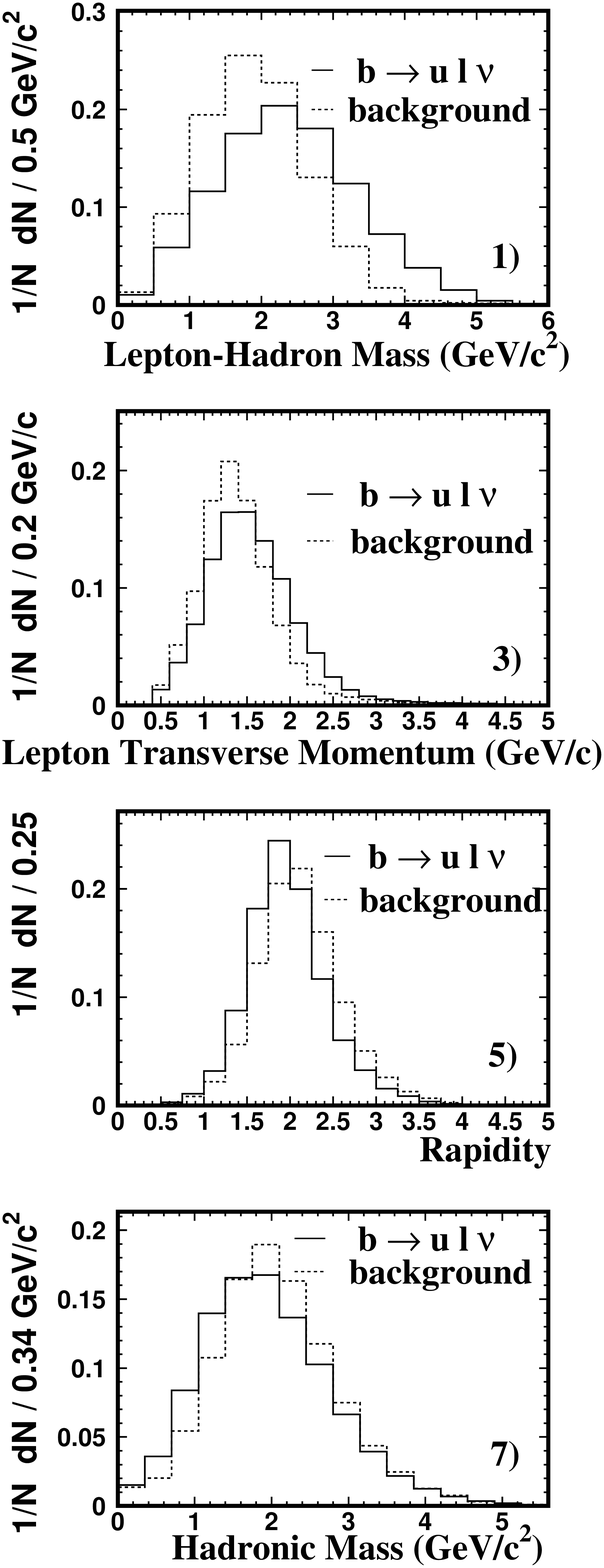,width=3.in}
\epsfig{file=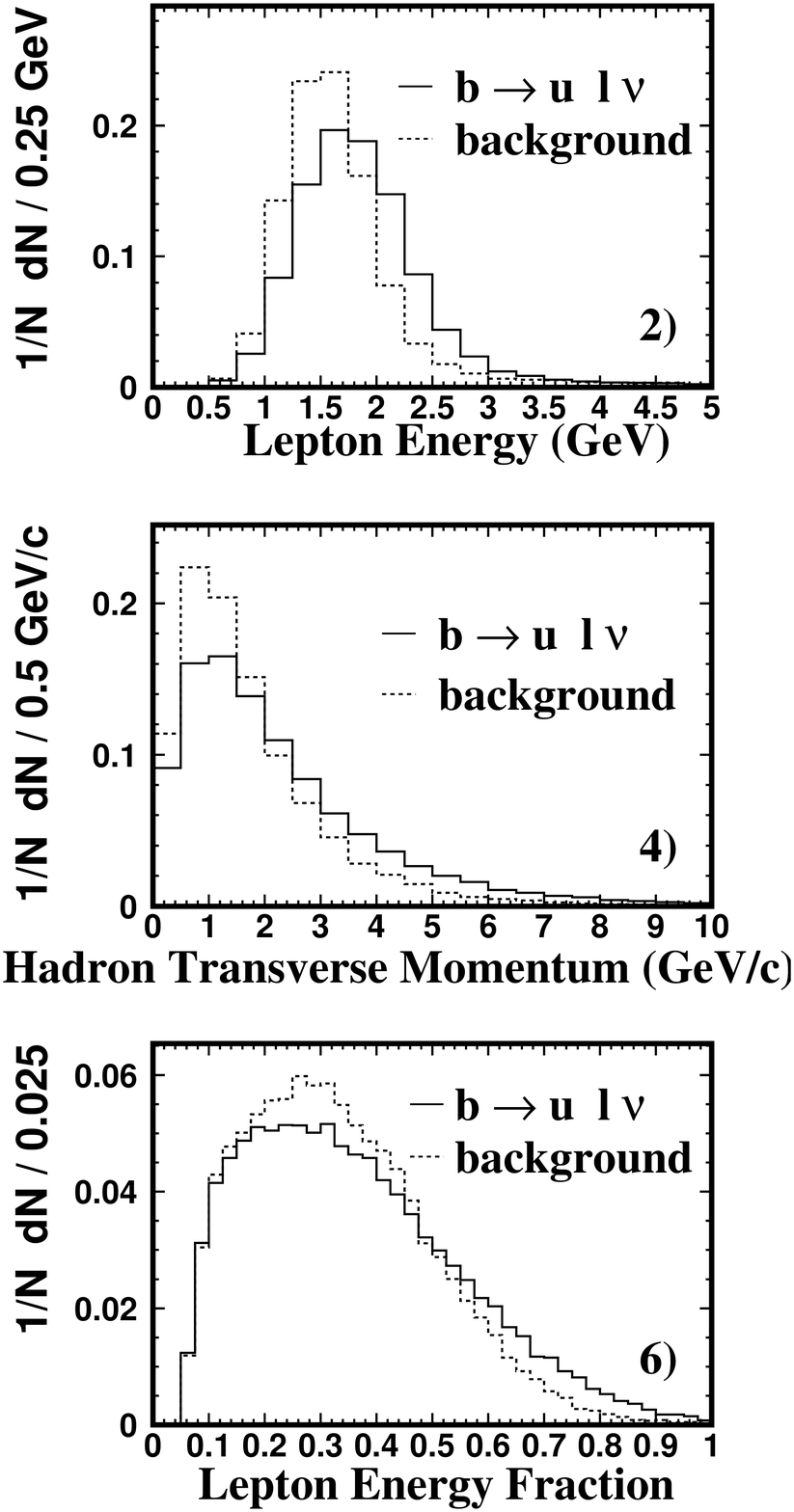,width=3.in}
\caption{\textbf{1-7.} Comparison  between the signal  \bu  and the
background  in  the Monte Carlo simulation for the seven 
\bu neural network input variables. 
The \bu signal and background are normalized to unity.
The input variables in plots 1 to 7 are in  the same order as  the input variables
defined  in the text of   Section~\ref{sec:5}.}
\label{fig:2302}
\end{figure}

\begin{figure}[htbp]
\centering
\centerline{\Large \textbf{OPAL}}
\epsfig{file=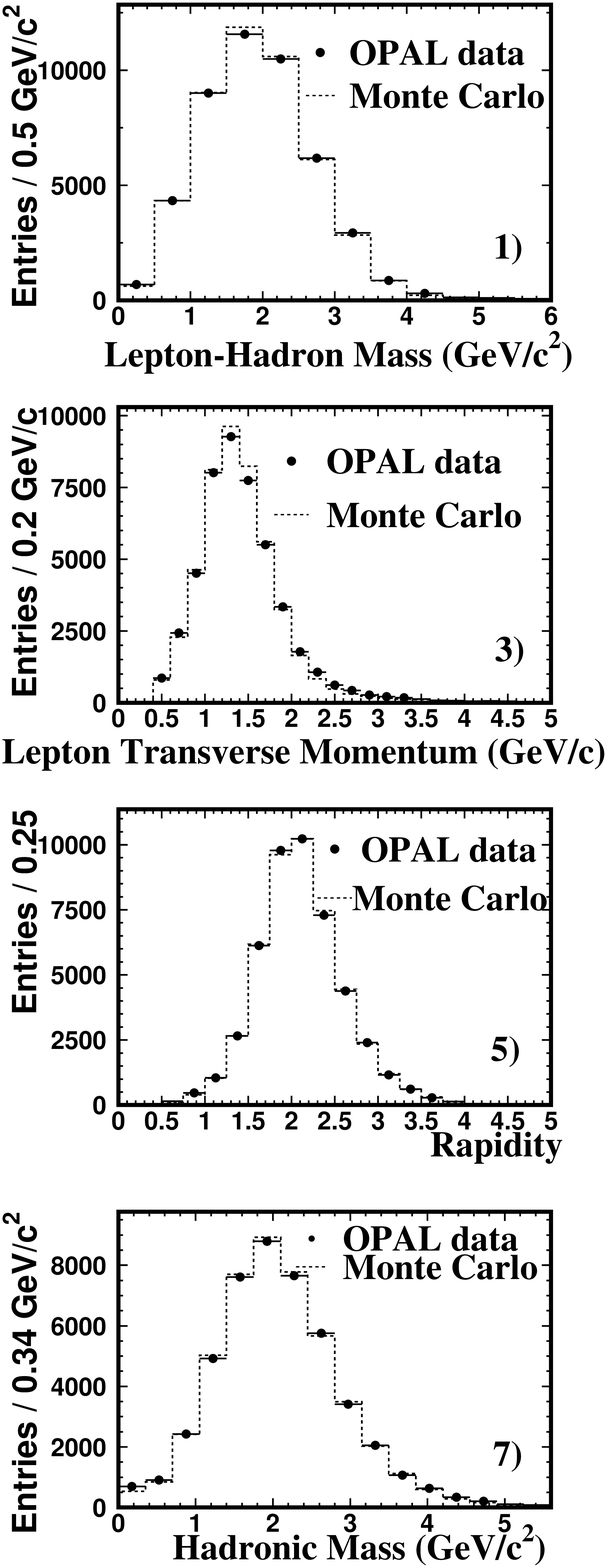,width=3.in}
\epsfig{file=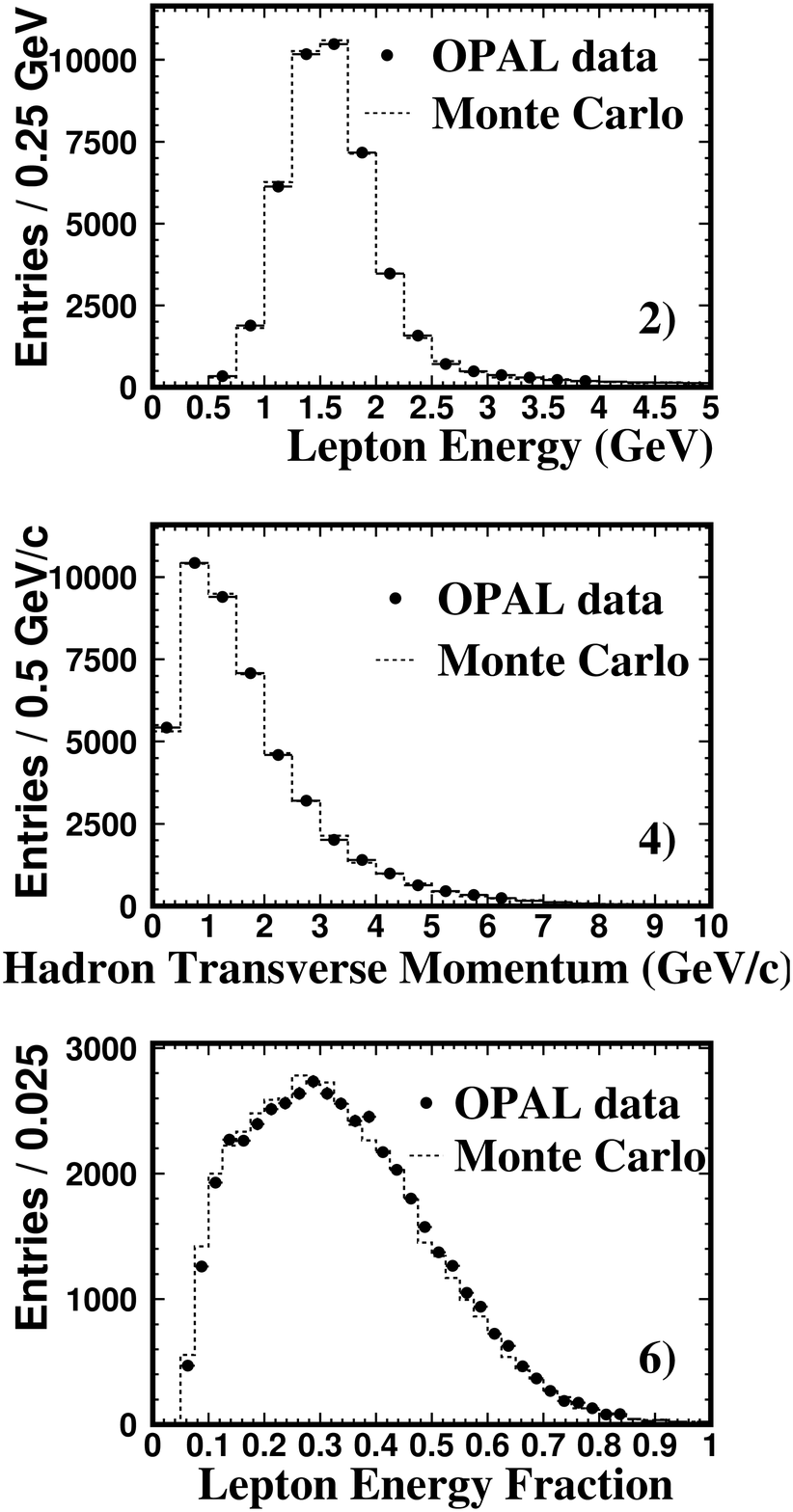,width=3.in}
\caption{\textbf{1-7.} \normalsize Comparison between the OPAL data   and the Monte
Carlo simulated events  for the seven neural network input
variables. A branching fraction of 1.63 $\times$ $10^{-3}$ for the \bu
transition is incorporated in the Monte Carlo simulation  for
comparison with  the OPAL data. The input
variables in plots 1 to 7 are in  the same order as  the input variables
defined  in the text of  Section~\ref{sec:5}.}
\label{fig:2320}
\end{figure}

Because of the dominant \bc background,
it is difficult to  enrich the sample in \bu decays 
 using only one kinematic variable.
A multi-layered feed-forward artificial neural network based on the JETNET 3.0
program\cite{jetnet} is used to enrich  the sample in \bu decays.
There are four layers in this neural network. The neural network
structure is 7-10-10-1. In the first layer,
seven variables are used as inputs to  the neural network.  
The last layer is the neural network output variable.
A figure of merit\cite{fig} is used  to determine 
the discrimination power of these seven variables  in separating  
two classes of events, i.e. signal and background.
The higher  the figure of merit, the better the separation between the two classes.
Over twenty kinematic variables were initially considered 
as inputs to the \bu neural network. Only seven variables are selected
as inputs to the \bu neural network based on good separation  between
\bu and background and good agreement between data and Monte Carlo
simulated events. These seven input variables,  in order
of decreasing figure of merit,  are:

 \begin{enumerate}
\item the invariant mass of the most energetic final state particle
combined with the lepton,
\item the lepton energy in the $b$ hadron rest frame, where the $b$ hadron
energy and momentum are estimated using the techniques described in
\cite{boost},
\item the lepton momentum transverse to the jet axis 
(the jet axis calculation includes the lepton),
\item  the transverse momentum of the most energetic final state 
hadron with respect to the lepton direction
(assuming all hadrons  are  pions),
\item  the rapidity of the most energetic final state hadron calculated
 with respect to the lepton direction (assuming all hadrons  are  pions),
\item the fraction of the reconstructed  $b$ hadron energy carried by the lepton,
\item the  reconstructed hadronic invariant mass, $\rm{M}_{\rm{x}}$, 
which is calculated by:
\begin{equation}
{\rm{M}_{\rm{x}}}^2 =  {\sum_{\rm{i}}(\rm{W}_{\rm{i}} \rm{E}_{\rm{i}})}^2-
{\ \sum_{\rm{i}} (\rm{W}_{\rm{i}} 
\vec{\rm{p}}_{\rm{i}})}^2,
\end{equation}
where i denotes all hadronic tracks and clusters. $\rm{W}_{\rm{i}}$ is 
the probability that the $\rm{i}^{\rm{th}}$ hadronic track or  unassociated  cluster  
comes from $b$ decay and is  calculated
using the techniques described in \cite{weight}. $\rm{E}_{\rm{i}}$ 
and $\vec{\rm{p}}_{\rm{i}}$ are the energy and momentum of the $\rm{i}^{\rm{th}}$ hadronic
track or neutral cluster.
\end{enumerate}
Only the tracks and clusters from the same jet as the lepton are
included in the calculation of these seven input variables.
The seven  input variable distributions for the \bu and the 
background in the Monte Carlo simulation are shown 
in Figure~\ref{fig:2302}.
The agreement between the Monte Carlo simulated events and the OPAL data for these  seven
variables is shown in Figure~\ref{fig:2320}.

\begin{figure}[htbp]
\centering
\epsfig{file=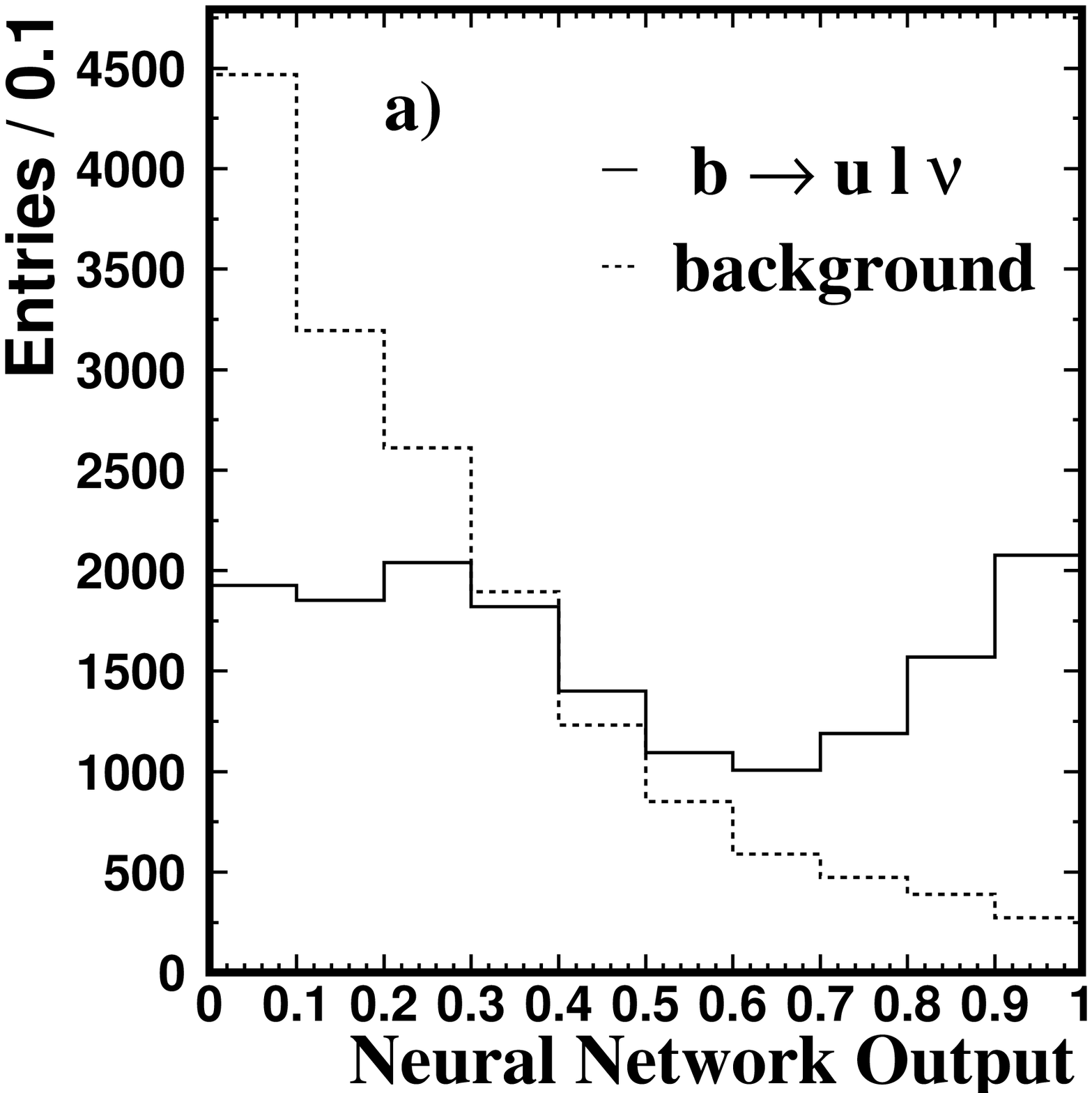 ,width=3. in}
\epsfig{file=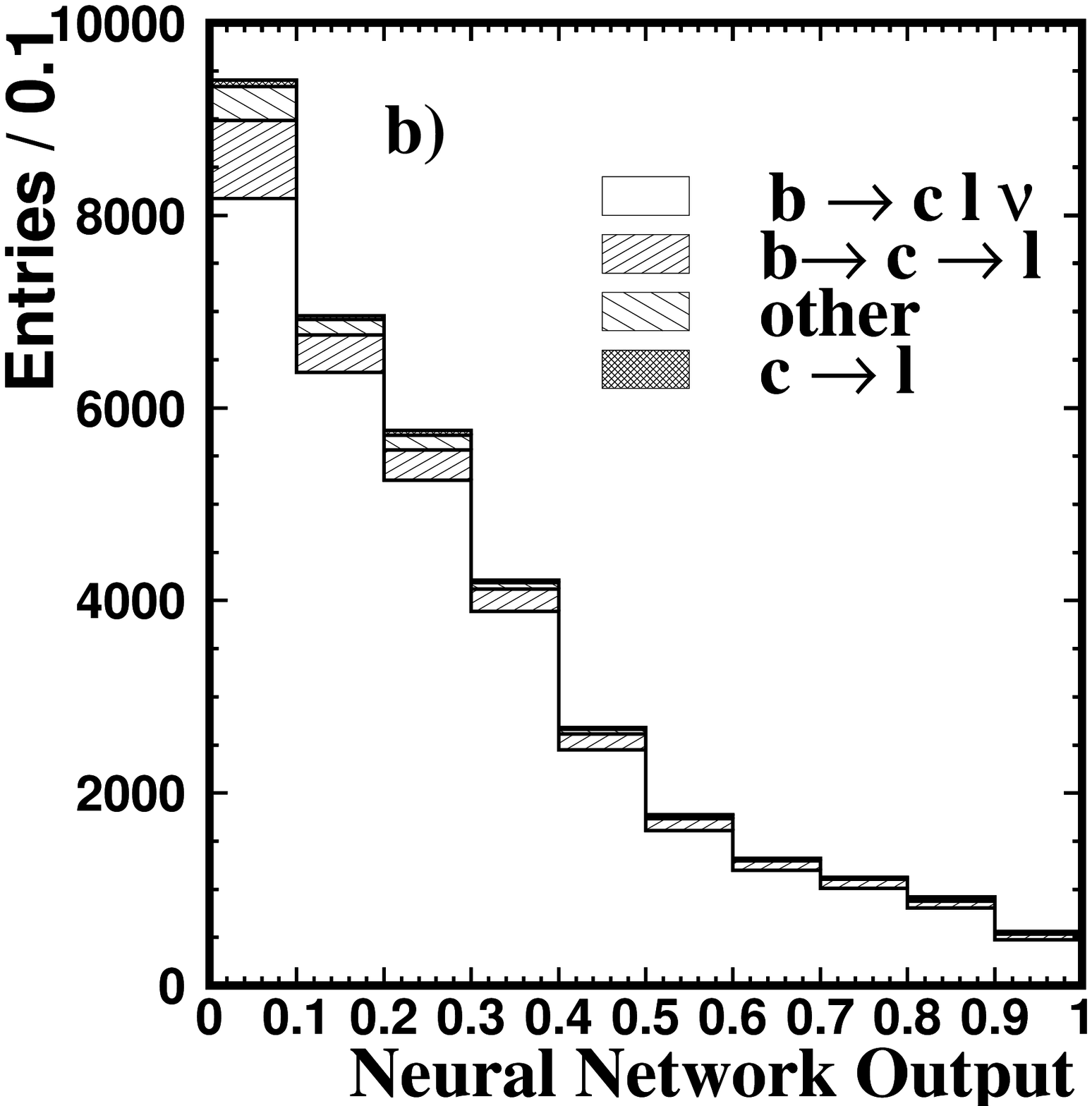 ,width=3. in}
\caption{\textbf{a,b.} The   \bu  neural network output distributions,
\textbf{a} for \bu and background, \textbf{b} for different
background components.}
\label{fig:training}
\end{figure}

Twelve thousand  \bu events, which  were
simulated with the hybrid model and have passed the event
preselection, and the same number of
background events from the  multi-hadron \( \rm{Z} \to \rm{q} 
\bar{\rm{q}}\)  Monte Carlo simulation
after the preselection are used  to train the \bu  neural network.
Two other samples of  signal and background
 events of the same size  are used to test the neural network performance.
The neural network output distributions  from   \bu and 
background  are shown in Figure \ref{fig:training}a.

The background composition from the  \bu neural
 network is shown in Figure~\ref{fig:training}b.
Ninety percent of the background in this analysis comes from  the 
\bc decay,  6.8$\%$ from  the \( \rm{b} \to \rm{c}\) decay with  the c subsequently 
decaying to a lepton. Another 0.6$\%$  comes
from the \(\rm{c} \to \ell\) decay in which the c quark is the 
primary quark. Other background processes make up the remaining 2.6$\%$, of which
36$\%$ is from the \( \rm{b} \to \tau \) decay with the $\tau$  subsequently
decaying to an electron or a muon, and  most of the rest of the  background is
from a  pion or a kaon misidentified as an electron or a muon.

\section{\boldmath Extraction of \brbu}
The branching fraction of  \bu decay can be
obtained from the  best fit of the Monte Carlo simulated events to OPAL
data  based on the \bu neural network output distributions.
\brbu is  extracted from the \bu neural network output distributions by
minimizing:
\begin{equation}
\rm{{\chi^2} = \sum_k {{[ N_k^{data}-N_{data}(\it{x}\ 
\rm{f_k^{MC_{bu}}} 
+ (1-\it{x})\rm{ f_k^{MC_{bg}}})]^2} \over {N_k^{data}}}},
\end{equation}
 where $\rm{N}_{\rm{k}}^{\rm{data}}$ is the number of events from the data in the 
$\rm{k}^{\rm{th}}$ bin of the neural network output. 
$\rm{N}_{\rm{data}}$ is the total
number of events in the data after preselection.
The  free parameter $x$ is
the fraction of signal events in the data after preselection, which can 
be converted to \brbu based on the number of signal events and the
number of background events in the 
Monte Carlo simulation after preselection.
$\rm{f}_{\rm{k}}^{\rm{MC}_{\rm{bu}}}$ is the  fraction  of simulated signal events 
 in the $\rm{k}^{\rm{th}}$ bin  of the \bu neural network output with
respect to the total number of simulated signal events after preselection.
$\rm{f}_{\rm{k}}^{\rm{MC}_{\rm{bg}}}$ is the  fraction  of simulated background events  
 in the $\rm{k}^{\rm{th}}$ bin of the \bu  neural network output 
with respect to the total number of simulated background events after preselection.
Here the background includes 
\(b \to \rm{X}_{\rm{c}} \ell \nu\), \(\rm{b} \to \rm{c} \to \ell\), 
\(\rm{c} \to \ell\) and other contributions.  
The sum over the index k is performed from the neural 
network cut to the last bin in the neural network output distribution. 
The  \(\brbr\) from the fit result $x$, as well as its statistical  and 
systematic errors, depends on the \bu  neural network cut.
The resulting \(\brbr\) is stable, with variations less than 0.1 $\times$ $10^{-3}$,
 as the neural network cut varies in  value from 0.3 to 0.7.
A neural network cut of 0.7 is chosen to minimize the total
relative errors and yields 
 \[\brbr  = (1.63 \pm 0.53) \times 10^{-3},\]
where the  uncertainty is the statistical error only.

\begin{figure}[htbp]
\centering
\epsfig{file=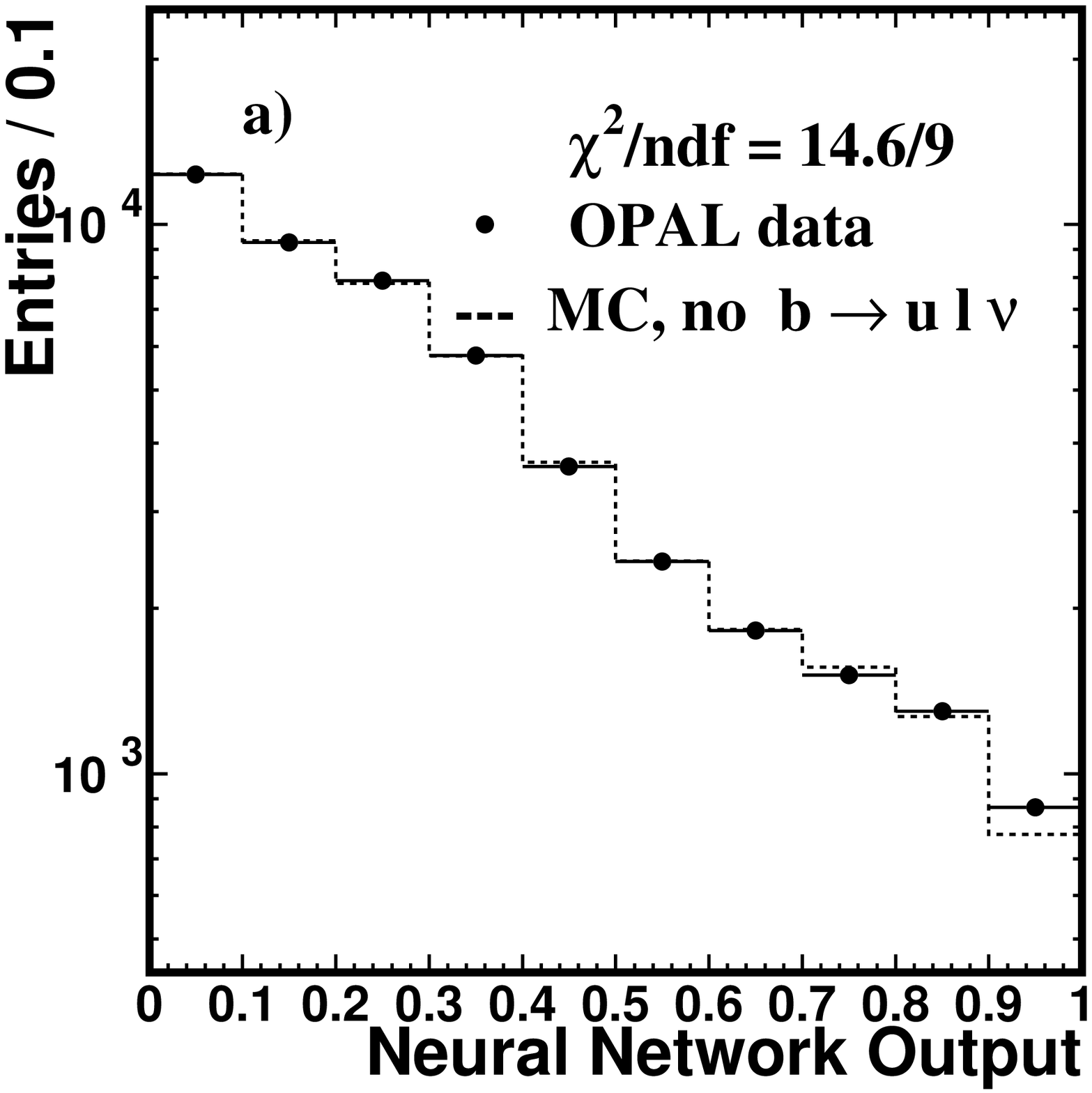,width=3. in}
\epsfig{file=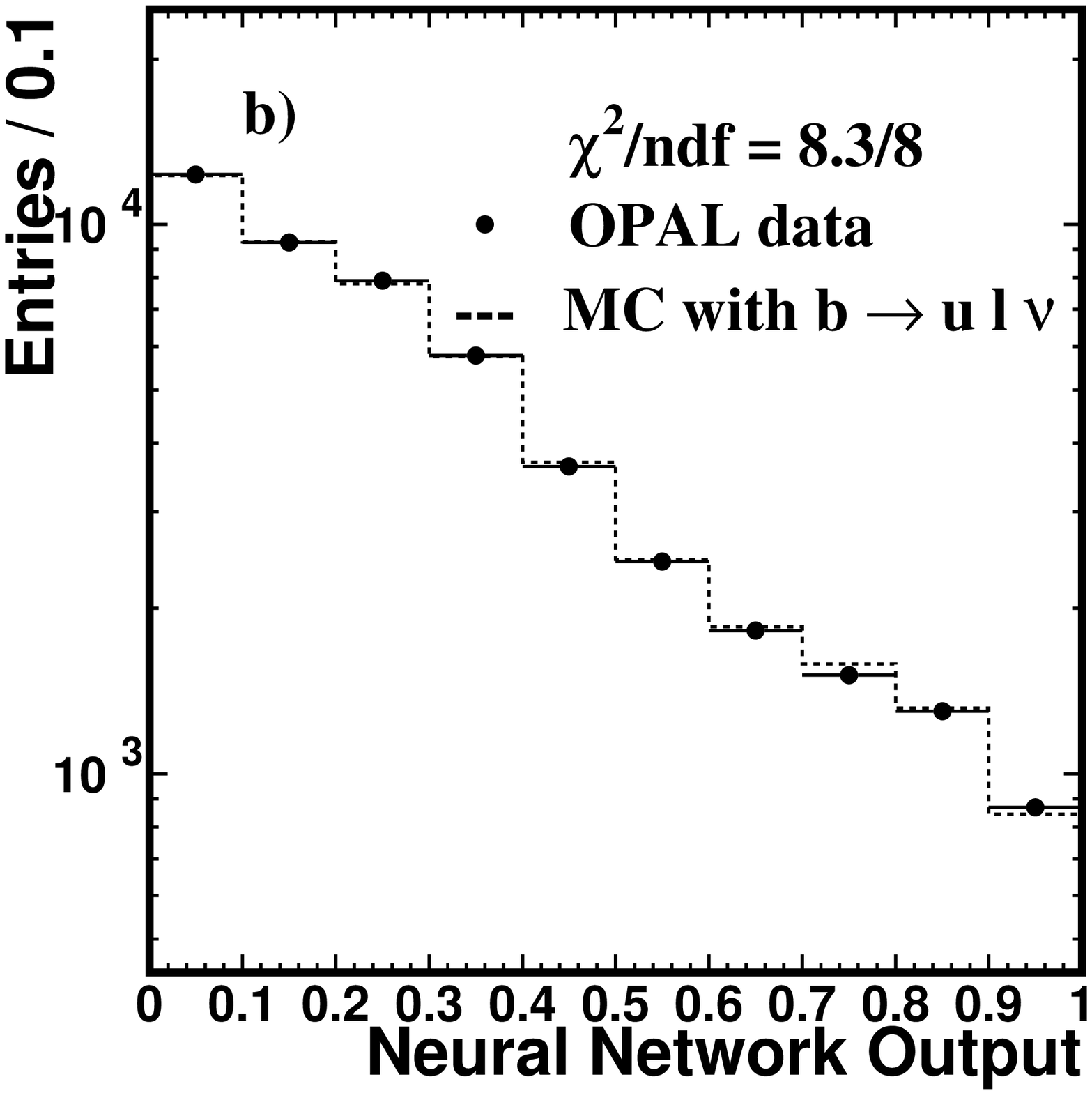,width=3. in }
\caption{\textbf{a,b.} The neural network output distributions for  data
and Monte Carlo  simulated events, \textbf{a}
with no \bu transition in the Monte Carlo simulated events, \textbf{b} 
 with a branching fraction of  1.63 $\times$ $10^{-3}$ \bu   decay  
incorporated. The distribution of Monte Carlo simulated events is
normalized to the data  for both plots.}
\label{fig:excess}
\end{figure}

In  Figure~\ref{fig:excess}a, the neural network output 
distribution from  data and the Monte Carlo simulation events  with
no \bu semileptonic decay is shown and the  excess of events in the
data can be seen in the last bin.
 Here the distribution of Monte Carlo  simulated events 
with no \bu transitions is
normalized to the same number of entries as the data.
The $\chi^2/\rm{ndf}$ is 14.6/9, which
corresponds to a  10$\%$ confidence level, assuming no
contributions from  \bu transition. Here the
$\chi^2$ is calculated by summing over all bins in the neural network 
output distribution. When the \bu transition   is
incorporated in the Monte Carlo simulation  with a branching
 fraction  of 1.63 $\times$ $10^{-3}$, 
the Monte Carlo simulation agrees much better with the data, as can be seen in
 Figure~\ref{fig:excess}b. The $\chi^2/\rm{ndf}$ is 
then 8.3/8, corresponding to a 41$\%$ confidence level.

\begin{figure}[htbp]
\centering
\epsfig{file=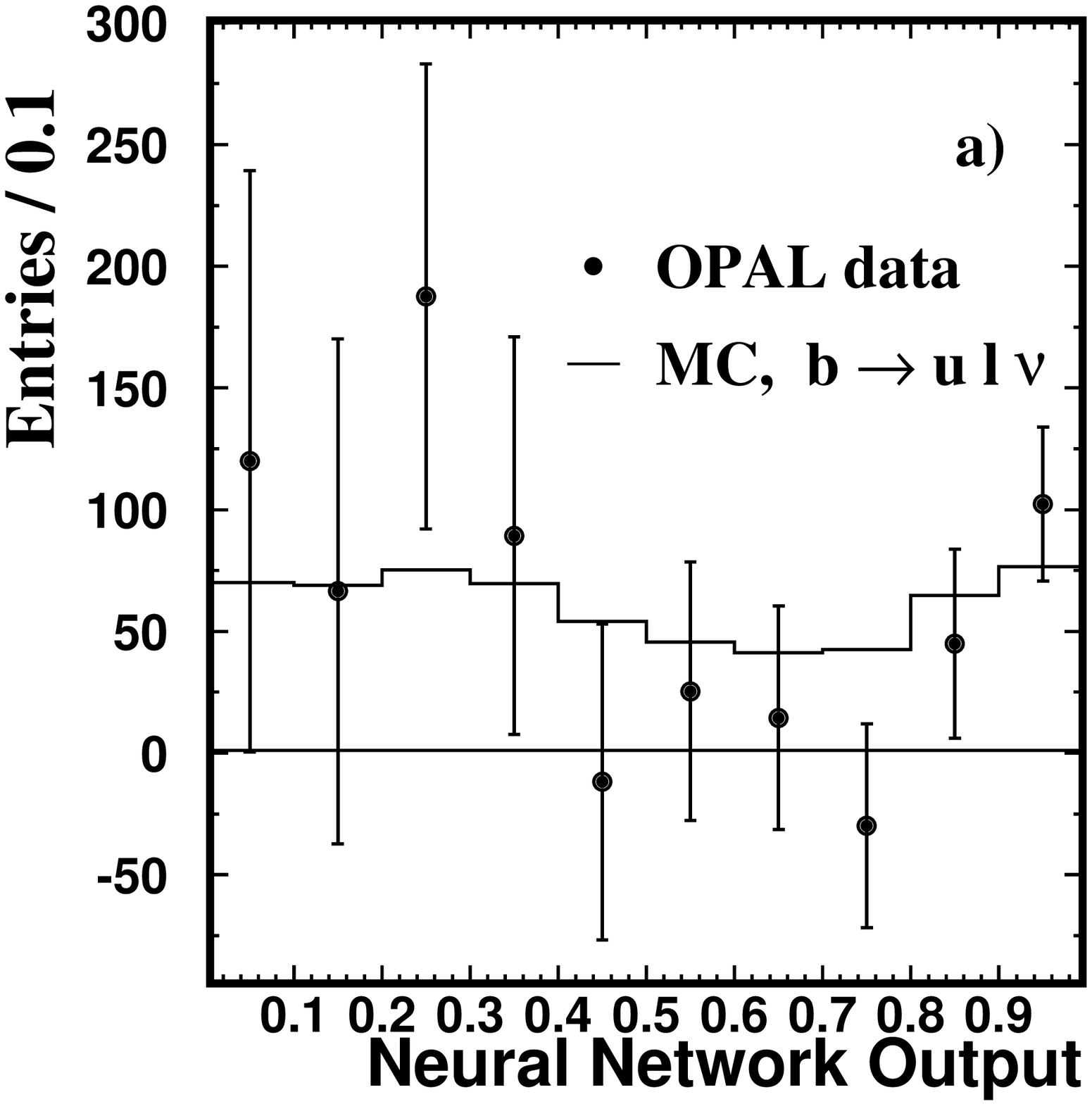,width=3. in}
\epsfig{file=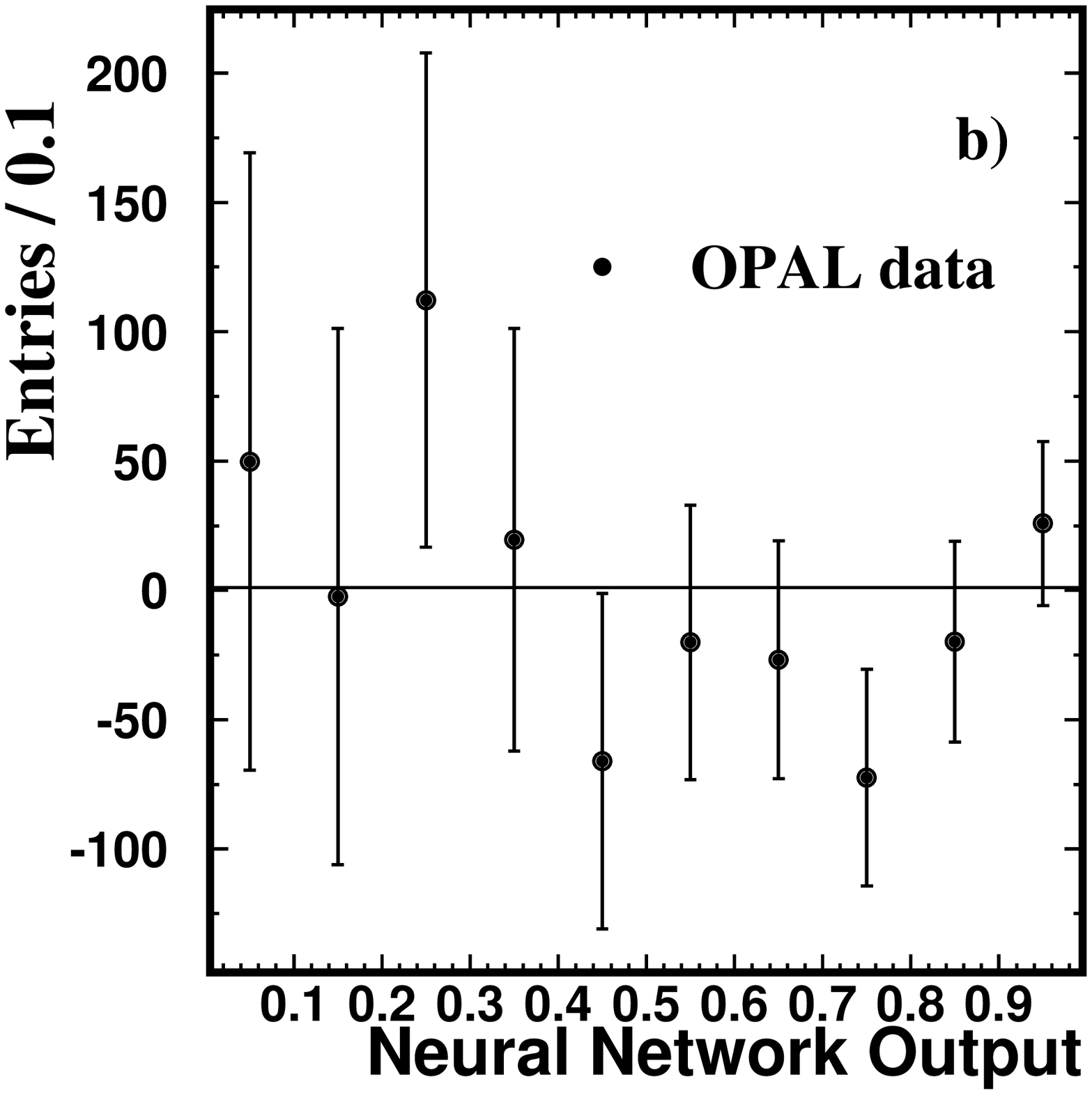,width=3. in }
\caption{\textbf{a,b.} The neural network output distributions. \textbf{a}
Data after subtracting the background from the Monte Carlo simulated events 
(points) show agreement with the simulated \bu signal (solid histogram).
\textbf{b} Data after subtracting the Monte Carlo simulated events 
 with a branching fraction of  1.63 $\times$ $10^{-3}$ \bu   decay  
incorporated. Here the error bars include the statistical error from
data  and Monte Carlo simulated events.}
\label{fig:excess1}
\end{figure}

The data after subtracting  the background from the Monte Carlo
simulated events agree well with the simulated \bu signal within 
statistical errors, which is shown in Figure~\ref{fig:excess1}.

A series of cross checks, dividing the lepton samples
into electron and
muon samples and dividing the data into two samples for the years
1991 to 1993 and the years  1994 to 1995,
are   performed.  
The \brbu results are consistent within statistical errors
for these cross checks.
\section{Systematic errors}
 The list of systematic errors is given in  Table \ref{table:2}.  
Unless otherwise specified, the systematic errors are estimated by
varying each parameter described by one standard deviation 
and taking the corresponding largest errors.
The resulting  systematic errors in  Table \ref{table:2}  are discussed in detail:
\begin{table}[htbp]
\centering
\begin{tabular}{|l | l| c|}
\hline  Error Source &  Variation or &  $\Delta$\brbu \\
                     &  value and variation        &  $10^{-3}$                            \\
\hline
Fragmentation  $\langle x_{\rm{E}} \rangle_{\rm{b}}$             & 
0.702 $\pm$ 0.008 \cite{xe}  &   $ ^{+0.28}_{-0.32}$\\
 Lepton spectrum (b $\to$ c)  &ISGW**\cite{CLEO1}, ISGW\cite{IGSW}&  $
^{+0.18}_{-0.29}$ \\
 MC statistics         & (see text)           & $\pm 0.22$      
\\
$b$ and $c$ hadron semileptonic decay     &  (see text)   &   $\pm 
0.19$ \\
MC modeling          &  (see text)    &   $\pm 0.19$ \\
\bu modeling error (hybrid)   & (see text)    &   $\pm 0.19$ \\
\bu modeling error (inclusive) & Parton\cite{parton}, QCD\cite{QCD}&  
$ \pm 0.14$ \\
\bu modeling error (exclusive) &  ISGW2\cite{ISGW2}, JETSET &  $\pm 0.07$ \\
Tracking resolution    &  $\pm$10$\%$\cite{rb}        & $\pm 0.07$     \\
 $c$ hadron decay multiplicity        &  (see text)             & $\pm 0.07$\\
 ${\Lambda}_{\rm{b}}$ production rate       & (11.6 $\pm$ 2.0)$\%$ 
\cite{PDG} &$ \mp 0.04$ \\
$\Lambda_{\rm{b}}$ polarization  & 
(${-56} ^{+43 }_{-31}$)$\%$\cite{polar} & $\pm 0.03$ \\
Electron ID efficiency & $\pm$4$\%$\cite{rb} & $\mp 0.04$\\
 Muon ID efficiency     & $\pm$2$\%$\cite{chris} & $\mp 0.03$ \\
Electron fake rate     & $\pm$21$\%$\cite{rb}     &  $\mp 0.02$ \\
Muon fake rate        &  $\pm$8$\%$\cite{rb}      &  $\mp 0.01$ \\
Br(\(b \to \rm{X} \tau \bar{\nu}_{\tau}\))           & (2.6 $\pm$ 
0.4)$\%$\cite{PDG}     & $\pm 0.01$ \\
$b$ lifetime              &(1.564 $\pm$ 0.014) ps\cite{PDG}  & $<$
0.01\\
$\rm{R}_{\rm{b}}$                    & 0.2178 $\pm$ 0.0017\cite{PDG}       
&$<$ 0.01 \\
\hline
 Total              &      &   $ ^{+0.55}_{-0.62}$ \\
\hline
\end{tabular} \\
\caption{Systematic errors for Br(\(b \to X_{u} \ell \nu\)).}
\label{table:2}
\end{table}

\begin{description}
\item[\boldmath b  quark fragmentation:]
Many parameterizations have been suggested to describe the heavy quark 
fragmentation process. The Peterson function\cite{Peterson}  is used
here to simulate the b and c fragmentation in the Monte Carlo simulation.
The systematic error in the b  quark fragmentation is estimated by
varying the $b$  hadron mean scaled energy $\langle x_{\rm{E}} \rangle_{\rm{b}}$ 
 within the experimental range 0.702 $\pm$ 0.008 recommended by the LEP
Electroweak Working Group \cite{xe}. This value is  consistent
with a recent determination of $\langle x_{\rm{E}} \rangle_{\rm{b}}$ = 
0.714 $\pm$ 0.009 from SLD \cite{sld1}. The systematic error is also estimated 
from  the  Collins and Spiller fragmentation function\cite{Collin}  and
Kartvelishvili fragmentation  function\cite{Kart}. The uncertainties
in  c quark fragmentation can be neglected because  the 
background from \( \rm{c} \to \ell \), where c is a primary quark, is
very small.

\item[\boldmath \bc lepton momentum spectrum modeling:]
Different decay models are used to predict the lepton spectrum in the
$b$  hadron rest frame for the \bc decay. Although all
models are derived for $\rm{B}^{0}$ and $\rm{B}^{+}$  semileptonic 
decay only, they are extrapolated to the
$\rm{B}_{\rm{s}}$ and $\Lambda_{\rm{b}}$ semileptonic decay. This 
will be correct in the simple spectator model
and is a reasonable  approximation for this analysis.
The lepton spectrum from  the ACCMM model\cite{ACCMM} is used as a base
model for the \( b \to \rm{X}_{\rm{c}} \ell \nu \) decay.
 The systematic errors due to \bc lepton
momentum spectrum modeling are estimated from the ISGW \cite{IGSW} and
ISGW**\cite{CLEO1} models as prescribed by the LEP Electroweak 
Working Group\cite{xe}.

The lepton spectrum from the $\rm{b} \to \rm{c} \to \ell$ decay in 
the ACCMM model is different  from the lepton spectrum in  the ISGW
model. The systematic error due to 
 the shape of the $\rm{b} \to \rm{c} \to \ell$ lepton spectrum  is 
calculated and is found to be negligible. The lepton spectrum from 
the  $\rm{c} \to \ell$ decay is varied  from the
ACCMM model to the ISGW model, where the c quark is a primary quark from Z
decay.  The systematic error is calculated and  found to be negligible.
\item[Monte Carlo statistics:]
The systematic uncertainty due to the limited Monte Carlo
statistics is $\pm$0.22$\times 10^{-3}$.
\item[\boldmath $b$ and $c$ hadron semileptonic decay branching 
fractions:]
The systematic error is estimated from the uncertainties of the  branching
fractions of
\( \rm{B} \to \rm{D} \ell \nu \), \( \rm{B} \to \rm{D}^* \ell \nu \),
\( \rm{B} \to \rm{D}^{**} \ell \nu\) and \( \Lambda_{\rm{b}} \to 
\Lambda_{\rm{c}} \rm{X}\ell\nu \).
There is a 6.8$\%$ background contribution from the \(\rm{b} \to 
\rm{c} \to \ell\) decays and a 0.6$\%$ background contribution  from the 
\(\rm{c} \to \ell\) decays. The systematic error is also estimated 
from the uncertainties of the branching fractions of the \( \rm{b} \to \rm{c} \to 
\ell\) decays. A summary of these systematic errors from the
uncertainties of $b$ hadron and $c$ hadron
semileptonic decay branching ratios is shown in
Table~\ref{table:3}. The Br\((\bar{\rm{B}} \to {\rm{D}}^{**} \ell
\nu)\) in Table~\ref{table:3} is obtained by averaging the
 Br\((\bar{\rm{B}} \to {\rm{D}}^{**} \ell\nu)\) from
ARGUS\cite{argus1}, ALEPH\cite{aleph1}, DELPHI\cite{delphi1}
and the total B  semileptonic decay branching fraction subtracting the 
contribution from  B to D and $\rm{D}^{*}$ semileptonic decay,  described
by the LEP, CDF and SLD Heavy Flavour Working Group \cite{lepwg}. 
For the decay of \(\bar{\rm{B}} \to {\rm{D}}^{**} \ell\nu\), in which
$\rm{D}^{**}$ refers to $\rm{D}_1$, $\rm{D}_2^{*}$, $\rm{D}_2$
and  $\rm{D}_1^{*}$, the branching ratio for each specific
$\rm{D}^{**}$ final state is not well measured.
For this analysis, the narrow final  states of $\rm{D}^{**}$
in \(\bar{\rm{B}} \to {\rm{D}}^{**} \ell\nu\)
are replaced by  the broad states and then vice-versa to check the
sensitivity of the \brbu to the relative ratio of the narrow and broad
states of $\rm{D}^{**}$ in \(\bar{\rm{B}} \to {\rm{D}}^{**} \ell\nu\).
The  effect on the  \brbu is found to be  negligible.

\begin{table}[htbp]
\centering
\begin{tabular}{|l | l| c|}
\hline  Error Source &  Variation &  $\Delta\brbr(10^{-3})$\\
\hline

Br\(( \rm{B}^{0} \to \rm{D}^{-}\ell^{+} \nu \)) & (2.10 $\pm$ 
0.19)$\%$\cite{PDG}  & $\mp$0.02 \\
Br\(( \rm{B}^{0} \to \rm{D}^{*-}\ell^{+} \nu \)) &  (4.60 $\pm$ 
0.27)$\%$\cite{PDG} & $\pm$0.03 \\
Br\( ({\rm{B}^{+}} \to \bar{\rm{D}}^{0}\ell^{+} \nu \)) & (2.15 $\pm$ 
0.22)$\%$\cite{PDG} & $\mp$0.06 \\
Br\(({\rm{B}^{+}} \to \bar{\rm{D}}^{*0}\ell^{+} \nu \)) & (5.3 $\pm$
0.8)$\%$\cite{PDG}  &  $\pm$0.04 \\
Br\((\bar{\rm{B}} \to {\rm{D}}^{**} \ell \nu)\) 
& (3.04 $\pm$ 0.44)$\%$ \cite{lepwg} & $\pm$0.16 \\
Br\( (\rm{b} \to  \rm{c} \to \ell) \) &   (8.4 $ 
^{+0.42}_{-0.39}$)$\%$\cite{chris}  & $\mp$0.02 \\
Br\( ( \Lambda_{\rm{b}} \to \Lambda_{\rm{c}} \rm{X}\ell\nu \)) & (7.9 
$\pm$1.9)$\%$\cite{PDG}  & $\mp$0.06          \\

\hline
 Total              &      &   $\pm$0.19 \\
\hline
\end{tabular} \\\
\caption{Systematic errors for \brbu from
uncertainties of the $b$
hadron and $c$ hadron semileptonic decay branching ratios.}
\label{table:3}
\end{table}
\item[Monte Carlo modeling errors:]
The systematic error for the Monte Carlo modeling errors  is estimated 
by re-weighting each input variable distribution in the Monte Carlo simulation
to agree with the corresponding data distributions. A
branching fraction of 1.63  $\times$ $10^{-3}$  for the \bu transition is
incorporated in the Monte Carlo simulation as shown in Figure 
\ref{fig:2320}. This gives a conservative
estimation of the systematic uncertainty due to the modeling of the input variables.
\item[\boldmath \bu modeling error from the
hybrid model:]
The boundary between the inclusive
and exclusive regions in the hybrid model is varied from 1.5 GeV/$c^2$
to 0.9 GeV/$c^2$. This conservatively estimates the systematic error 
arising from the placement of the boundary between
the inclusive  and  exclusive models.
This produces a uncertainty of  $\pm$0.19$\times 10^{-3}$ for 
Br(\(b \to \rm{X}_{\rm{u}} \ell \nu\)).
\item[\boldmath \bu inclusive model:]
The ACCMM model is the base inclusive model. The  QCD universal
function model and the parton model are used to estimate the systematic errors in the
inclusive part of the \bu hybrid  model. This gives a change of -0.14
$\times$ $10^{-3}$ for the QCD universal function model and +0.02 $\times$
$10^{-3}$  for the parton model for the branching ratio of 
\(b \to \rm{X}_{\rm{u}} \ell \nu \). The largest variation of \brbu
from these models is taken as the
systematic uncertainty of \brbu from the inclusive model.
\item[\boldmath \bu exclusive model:]
The ISGW2 model is the base exclusive model.
The model implemented in
the OPAL tune of JETSET\cite{para} Monte Carlo simulation, 
which has the u quark and the spectator quark forming one single hadron in
the final state,  is used to estimate the systematic error in the
exclusive part of the \bu hybrid  model. 
\item[Tracking resolution:]
The systematic error due to the uncertainties of the detector
resolution is estimated by applying a $\pm 10\%$  variation 
to the $r$-$\phi$ track parameters and an independent $\pm 10\%$
variation  to the analogous parameters in the $r$-$z$  plane  
to the Monte Carlo simulated events \cite{rb}. 
\item[\boldmath $c$  hadron decay multiplicity:]
The systematic error of the
\brbu associated with the $c$  hadron decay 
charge multiplicity is estimated using the
average charged track multiplicity of $\rm{D}^{+}$, $\rm{D}^{0}$, 
$\rm{D}_{\rm{s}}^{+}$
decays as measured by MARK III  \cite{multip}. The systematic uncertainty
of the \brbu is $\pm$0.07$\times 10^{-3}$ from the uncertainty
of $c$  hadron decay multiplicity.
\item[\boldmath ${\Lambda}_{\rm{b}}$ production rate:]
The PDG\cite{PDG}  gives the production fraction of
 $\rm{B}^{+}$, $\rm{B}^0$, $\rm{B}^0_s$ and  $\Lambda_{\rm{b}}$
in Z decay as (38.9 $\pm$ 1.3)$\%$, (38.9 $\pm$ 1.3)$\%$,
(10.7 $\pm$ 1.4)$\%$ and (11.6 $\pm$ 2.0)$\%$ respectively.
 The neural network output variable distributions among  $\rm{B}^{+}$,
  $\rm{B}^0$ and $\rm{B}^0_{\rm{s}}$ are similar and the systematic 
effects caused by the uncertainties of the  production
fraction of $\rm{B}^{+}$,  $\rm{B}^0$ and $\rm{B}^0_{\rm{s}}$ are 
negligible. Due  to the difference of the  neural network output variable 
distributions  between $\Lambda_{\rm{b}}$ and B mesons,
the fraction of  $\Lambda_{\rm{b}}$ is varied by  one standard 
deviation  to determine the corresponding
systematic error.
\item[\boldmath $\Lambda_{\rm{b}}$ polarization:]
The  lepton momentum spectrum from $\Lambda_{\rm{b}}$ semileptonic 
decays depends on the degree of  $\Lambda_{\rm{b}}$ polarization.
The systematic uncertainties are estimated by  using the 
$\Lambda_{\rm{b}}$ polarization ranging from  -13$\%$  to -87$\%$, as 95$\%$
\cite{polar} confidence level limits,
 which are converted to  one standard  deviation in
Table~\ref{table:2}.
\item[Lepton identification  efficiency:]
The number of selected  events in the signal and background depends
on the electron identification efficiency and the muon identification  efficiency.
The electron identification efficiency has been studied using control
samples of electrons from \( \rm{e}^{+}\rm{e}^{-} \to 
\rm{e}^{+}\rm{e}^{-} \) events and photon conversions, and is modeled to a precision
of 4$\%$\cite{rb}. The muon identification efficiency has been studied
by using the muon pairs produced in two-photon collisions and
\( \rm{Z} \to \mu^{+}\mu^{-}\)  events, giving an uncertainty of 
2$\%$ \cite{chris}.
\item[Lepton fake rate:]
Fake electrons in the  electron sample  are primarily  from  charged
hadrons (mainly charged pions) misidentified as electrons and from untagged
photon conversions.  The uncertainty associated with  electron misidentification
is $\pm$21$\%$\cite{rb}. The muon fake rate is studied from 
\( \rm{K}^0_{\rm{s}} \to \pi^{+} \pi^{-} \) and \( \tau \to 3\pi\)
decay. The uncertainty of the fake muon rate  is estimated to be $\pm$8$\%$.  
\item[\boldmath \(b \to X \tau  \bar{\nu}_{\tau}\)  branching 
ratio:]
One important  composition in the ``other'' background in
Figure~\ref{fig:training}b results from   a  b  quark semileptonic decay  to a $\tau$ lepton,
with the $\tau$ lepton subsequently decaying to an electron or a
muon.  The  branching ratio of  \(b \to X \tau  \bar{\nu}_{\tau}\)
is  (2.6 $\pm$ 0.4)$\%$\cite{PDG}. The systematic error
is estimated using the uncertainties of the  \( b \to \rm{X} \tau 
\bar{\nu}_{\tau}\) branching ratio.
\item[\boldmath Uncertainty of the $b$ lifetime]
The  average $b$ hadron lifetime is measured to be (1.564 $\pm$ 0.014)~ps
\cite{PDG}. The uncertainty in  $b$ lifetime  results in a negligible
uncertainty in Br(\(b \to \rm{X}_{\rm{u}} \ell \nu\)).
\item[\boldmath Uncertainty of $\rm{R}_{\rm{b}}$:]
The fraction of \(\rm{Z} \to \rm{b} \bar{\rm{b}} \) events in  
hadronic Z decays, $\rm{R}_{\rm{b}}$, is measured to be 0.2178 $\pm$
0.0017\cite{PDG}. The uncertainty in   $\rm{R}_{\rm{b}}$ results in a
negligible uncertainty in the background composition.

\end{description}

\section{Conclusion}
The branching fraction of the inclusive \bu decay  is measured to be:
 \[ \brbr = (1.63 \pm 0.53\ (\rm{stat})\   ^{+0.55}_{-0.62}\ 
(\rm{sys})) \times 10^{-3}.\]
The first error 0.53 is the statistical error from the data only.
The errors  associated with the limited statistics of  the Monte Carlo sample
are included in the systematic error.
This  result is consistent  with  similar  measurements
from   ALEPH, DELPHI and  L3, the other three LEP 
experiments, shown in Table \ref{table:result}. In  Table \ref{table:result},
the first error in  \brbu
combines the statistical error  from the data and  limited Monte Carlo
 statistics as well as  the  uncorrelated
systematic uncertainties due to   experimental systematic errors, 
such as detector resolution and lepton identification
efficiency. The second error contains the systematic uncertainties 
from the \bc Monte Carlo simulation models. The third error  contains
the systematic uncertainties from the \bu models.  The second and
third errors are correlated between the various experiments. \\
\begin{table}[htbp]
\begin{tabular}{|l | l| l | }
\hline  Experiment & \brbu ($10^{-3}$) & Ref \\

\hline
           ALEPH   &1.73 $\pm$ 0.56 (stat+det) $\pm$ 0.51 (b $\to$ c)
           $\pm$ 0.2 (b $\to$ u) & \cite{ALEPH} \\
\hline
           DELPHI  &1.69 $\pm$ 0.53 (stat+det) $\pm$ 0.45 (b $\to$
           c) $\pm$ 0.2 (b $\to$ u) & \cite{DELPHI}\\
\hline
            L3     &3.3 $\pm$ 1.3 (stat+det) $\pm$ 1.4 (b $\to$ c)
$\pm$ 0.5 (b $\to$ u) & \cite{L3} \\
\hline
           OPAL (This analysis) &1.63 $\pm$ 0.57 (stat+det)  $
^{+0.44}_{-0.52} $ (b $\to$ c) $\pm$ 0.25 (b $\to$ u) & \\
 \hline
\end{tabular} \\\
\caption{ \brbu results from  ALEPH, DELPHI, L3 and this analysis.}
\label{table:result}
\end{table}  \\
\ub can be obtained from \brbu
\cite{vub,vub.1} with inputs slightly revised as described by the LEP
Heavy Flavour
Working Group \cite{lepwg} in the context of
the Heavy Quark Expansion\cite{HQE}:
\begin{equation}
 \uub = 0.00445 \times \left( {{\brbr} \over 0.002} 
{1.55\rm{ps} \over \tau_{\rm{b}}}\right)^{1 \over 2}\times (1 \pm
0.010_{\rm{pert}} \pm 0.030_{1/\rm{m}_{\rm{b}}^{3}} \pm 
0.035_{\rm{m}_{\rm{b}}}), 
\end{equation}
where the average $b$ hadron lifetime $\tau_{\rm{b}}$ is equal to  
(1.564 $\pm$ 0.014) ps\cite{PDG} from the LEP
experiments.  Thus, \ub obtained from this analysis is: 
\[\uub = (4.00 \pm  0.65\ (\rm{stat})\    ^{+0.67}_{-0.76}\ (\rm{sys}) 
\pm 0.19\ (\rm{HQE})) \times 10^{-3},\]
where the systematic error includes the b to u and b to c  semileptonic
decay modeling error, and the HQE error is purely the theoretical
error from the Heavy Quark Expansion.
This result is consistent with the   \ub  value
from the CLEO exclusive measurement of 
(3.3 $\pm$ 0.8\ (total)) $\times$  $10^{-3}$\cite{CLEO3}.

\section*{Acknowledgments}
We particularly wish to thank the SL Division for the efficient operation
of the LEP accelerator at all energies
 and for their continuing close cooperation with
our experimental group.  We thank our colleagues from CEA, DAPNIA/SPP,
CE-Saclay for their efforts over the years on the time-of-flight and
trigger systems which we continue to use.  In addition to the support
staff at our own
institutions we are pleased to acknowledge the  \\
Department of Energy, USA, \\
National Science Foundation, USA, \\
Particle Physics and Astronomy Research Council, UK, \\
Natural Sciences and Engineering Research Council, Canada, \\
Israel Science Foundation, administered by the Israel
Academy of Science and Humanities, \\
Minerva Gesellschaft, \\
Benoziyo Center for High Energy Physics,\\
Japanese Ministry of Education, Science and Culture (the
Monbusho) and a grant under the Monbusho International
Science Research Program,\\
Japanese Society for the Promotion of Science (JSPS),\\
German Israeli Bi-national Science Foundation (GIF), \\
Bundesministerium f\"ur Bildung und Forschung, Germany, \\
National Research Council of Canada, \\
Research Corporation, USA,\\
Hungarian Foundation for Scientific Research, OTKA T-029328,
T023793 and OTKA F-023259.\\



\newpage
\begin{thebibliography}{99}
\bibitem{CKM}
N.\ Cabibbo,   Phys. Rev. Lett. \textbf{10},  531 (1963);
M.\ Kobayashi and K.\ Maskawa, 
Prog. Theor. Phys. \textbf{49},  652 (1973)


\bibitem{ARGUS}
ARGUS Collaboration, H.\ Albrecht \etal, Phys. Lett. \textbf{B234},  409  (1990)

\bibitem{CLEO4}
CLEO Collaboration, R.\ Fulton \etal, Phys. Rev. Lett. \textbf{64}, 16 
(1990);
CLEO Collaboration, J.\ Bartelt,  Phys. Rev. Lett. \textbf{71}, 
4111 (1993)


\bibitem{ALEPH}
ALEPH Collaboration, R.\ Barate \etal,
Eur. Phys. J. \textbf{C6},  555 (1999)

\bibitem{DELPHI}
DELPHI Collaboration, P.\ Abreu \etal,
Phys. Lett. \textbf{B478},  14  (2000)

\bibitem{L3}
L3 Collaboration, M.\ Acciarri \etal,
Phys. Lett. \textbf{B436}, 174 (1998)


\bibitem{HQE}
A.H.\ Hoang, Z.\ Ligeti and A.V.\ Manohar, 
Phys. Rev. \textbf {D59},  074017  (1999);
A.H.\ Hoang, Z.\ Ligeti and A.V.\ Manohar, 
Phys. Rev. Lett. \textbf{82},  277 (1999)

\bibitem{CLEO99}
CLEO Collaboration, B.H.\ Behrens \etal, 
 Phys. Rev. \textbf{D61},  052001 (2000)

\bibitem{detector}
OPAL Collaboration, K.\ Ahmet \etal, 
Nucl. Instrum. Methods \textbf{A305}, 275 (1991);
P.P.\ Allport \etal, 
Nucl. Instrum. Methods \textbf{A324}, 34 (1993);
P.P.\ Allport \etal,   Nucl. Instrum. Methods \textbf{A346},  476 (1994) 

\bibitem{JETSET}
T.\ Sj{\"o}strand, 
 Comp. Phys. Comm. \textbf{82},  74 (1994)


\bibitem{para}
OPAL Collaboration, G.\ Alexander \etal,  Z. Phys. \textbf{C69}, 543 (1996)

\bibitem{hybrid}
C.\ Ramirez, J.F.\ Donoghue and G.\ Burdman,  Phys. Rev.
\textbf{D41}, 1496 (1990)


\bibitem{gopal}
J.\ Allison \etal,  Nucl. Instrum. Meth. \textbf{A317}, 47 (1992)

\bibitem{PDG}
D.E.\ Groom \etal, Eur. Phys. J. \textbf{C15},  1 (2000)

\bibitem{bound1}
B.\ Grinstein, M.B.\ Wise, and N.\ Isgur,
Phys. Rev. Lett. \textbf{56}, 298 (1986)

\bibitem{IGSW}
N.\ Isgur, D.\ Scora, B.\ Grinstein and M.\ B.\ Wise,  Phys.
Rev. \textbf{D39},  799  (1989)

\bibitem{bound2}
M.\ Wirbel, B.\ Stech and M.\ Bauer, 
Z. Phys. \textbf{C29}, 637 (1985);
J.G.\ K{\"o}rner and G.A.\ Schuler, 
Z. Phys. \textbf{C38},  511 (1988)

\bibitem{ISGW2}
D.\ Scora and N.\ Isgur,  Phys. Rev.
\textbf{D52}, 2783  (1995)

\bibitem{ACCMM}
G.\ Altarelli, N.\ Cabibbo, G.\ Corb\`o, L.\ Maiani and
G.\ Martinelli, Nucl. Phys. \textbf{B208},  365 (1982)


\bibitem{QCD}
T.\ Mannel and  M.\ Neubert,  Phys. Rev. \textbf{D50},  2037 (1994)
\bibitem{QCD1}
M.\ Neubert,  Phys. Rev. \textbf{D49}, 4623 (1994)

\bibitem{QCD3}
R.D.\ Dikeman, M.\ Shifman and N.G.\ Uraltsev,
Int. J. Mod. Phys. \textbf{A11},  571 (1996)

\bibitem{parton}
A.\ Bareiss and E.A.\ Paschos, Nucl. Phys.
\textbf{B327},  353 (1989)

\bibitem{wdecay}
M.\ Jezabek and J.H.\ K$\mathrm{\ddot{u}}$hn,
Nucl. Phys. \textbf{B314}, 1 (1989) 


\bibitem{Peterson}
C.\ Peterson, D.\ Schlatter, I.\ Schmitt and P.\ M.\ Zerwas,
Phys. Rev. \textbf{D27},  105 (1983)


\bibitem{polar}
OPAL Collaboration, G.\ Abbiendi \etal,
Phys. Lett. \textbf{B444},  539 (1998)


\bibitem{electron}
OPAL Collaboration, R.\ Akers \etal, 
Z. Phys. \textbf{C66},  555 (1995)


\bibitem{rb}
OPAL Collaboration, G.\ Abbiendi \etal, Eur. Phys. J. \textbf{C8},
217 (1999)
\bibitem{cone}
OPAL Collaboration, R.\ Akers \etal,
Z. Phys. \textbf{C63},  197 (1994)

\bibitem{chris}
OPAL Collaboration, G.\ Abbiendi \etal, 
Eur. Phys. J. \textbf{C13},  225 (2000)

\bibitem{jetnet}
C.\ Peterson, T.\ Rognvaldsson and L.\ Lonnblad,  Comp. Phys. Comm.
\textbf{81},  185 (1994)

\bibitem{fig}
G.\ Bahan and R.\ Barlow, Comp. Phys. Comm. \textbf{74}, 199 (1993)

\bibitem{boost}
OPAL  Collaboration,  R.\ Akers \etal,  Z. Phys. \textbf{C66}, 19 (1995)


\bibitem{weight}
OPAL Collaboration, K.\ Ackerstaff  \etal,
Z. Phys. \textbf{C74},  413 (1997)

\bibitem{xe}
The LEP collaborations, ALEPH, DELPHI, L3 and OPAL,
 Nucl.\ Instrum. Methods \textbf{A378},  101 (1996)



\bibitem{CLEO1}
CLEO Collaboration, R.\ Fulton \etal,  Phys. Rev. \textbf{D43},
651 (1991)


\bibitem{sld1}
SLD collaboration, K.\ Abe \etal, Phys. Rev. Lett. \textbf{84}, 4300 (2000)
\bibitem{Collin}
P.D.B.\ Collins and T.P.\ Spiller,  J. Phys. \textbf{G11}, 1289 (1985) 

\bibitem{Kart}
V.G.\ Kartvelishvili, A.K.\ Likhoded and V.A.\ Petrov, 
Phys. Lett. \textbf{B78},  615  (1978)

\bibitem{argus1}
ARGUS Collaboration, H.\ Albrecht \etal, Z. Phys. \textbf{C57},  533 (1993)

\bibitem{aleph1}
ALEPH Collaboration, D.\ Buskulic \etal, Z. Phys. \textbf{C73}, 601 (1997)

\bibitem{delphi1}
DELPHI Collaboration, P.\ Abreu \etal, 
Phys. Lett. \textbf{B475},  407 (2000)


\bibitem{lepwg}
ALEPH Collaboration, CDF Collaboration,  DELPHI Collaboration,  L3
Collaboration, OPAL Collaboration and SLD Collaborations,  D.\ Abbaneo
\etal, ``Combined results on B hadron production rates, lifetimes,
oscillations and semileptonic decays'', SLAC-PUB-8492,
CERN-EP-2000-096  (2000)

\bibitem{multip}
MARK III  Collaboration, D.\ Coffman \etal, 
Phys. Lett. \textbf{B263},  135 (1991)

\bibitem{vub}
I.\ Bigi, R.D.\ Dikeman, N.\ Uraltsev \etal, 
Eur. Phys. J. \textbf{C4},  453  (1998)

\bibitem{vub.1}
N.\ Uraltsev, 
Int. J. Mod. Phys. \textbf{A14},  4641  (1999)

\bibitem{CLEO3}
CLEO Collaboration, J.P.\ Alexander \etal,
Phys. Rev. Lett. \textbf{77},  5000 (1996) 

\end{thebibliography}
\end{document}